\documentstyle[prb,aps,epsf,multicol,amssymb]{revtex}
\draft
\begin{document}

%
\newcommand{\fig}[2]{\epsfxsize=#1\epsfbox{#2}}
%
%
%
 \newcommand{\passage}{
         \end{multicols}\widetext\noindent\rule{8.8cm}{.1mm}\rule{.1mm}{.4cm}} 
 \newcommand{\retour}{
         \hspace{1cm}
         \noindent\rule{9.1cm}{0mm}\rule{.1mm}{.4cm}\rule[.4cm]{8.8cm}{.1mm}
         \begin{multicols}{2} }
 \newcommand{\unecol}{\end{multicols}}
 \newcommand{\deuxcol}{\begin{multicols}{2}}
%
%

\title{Glass phase of two-dimensional triangular elastic lattices with disorder}

\author{David Carpentier and Pierre Le Doussal}
\address{CNRS-Laboratoire de Physique Th\'eorique de l'Ecole\\
Normale Sup\'erieure, 24 rue Lhomond, 75231 Paris
Cedex-France\thanks{LPTENS is a Unit\'e Propre du C.N.R.S. 
associ\'ee \`a l'Ecole Normale Sup\'erieure et \`a l'Universit\'e Paris Sud}}

\maketitle

\begin{abstract}
We study two dimensional triangular elastic lattices in
a background of point disorder, excluding dislocations (tethered
network). Using both (replica symmetric) static and (equilibrium) dynamic
renormalization group for the corresponding $N=2$ component model,
we find a transition 
to a glass phase for $T < T_g$, described by {\it a plane} of perturbative
fixed points. The growth of displacements is found to be asymptotically
isotropic with $u_T^2 \sim u_L^2 \sim A_1 \ln^2 r$, with
universal subdominant anisotropy $u_T^2 - u_L^2 \sim A_2 \ln r$.
where $A_1$ and $A_2$ depend continuously
on temperature and the Poisson ratio $\sigma$. We also obtain
the continuously varying dynamical exponent $z$. For the
Cardy-Ostlund $N=1$ model, a particular case of the
above model, we point out a discrepancy in the value of $A_1$
with other published results in the litterature. We find that
our result reconciles the order of magnitude of the RG predictions
with the most recent numerical simulations.
\end{abstract}
\pacs{LPTENS preprint 96/065}

\deuxcol

\section{Introduction}

There is a lot of current interest in the problem
of elastic lattices in the presence of substrate disorder.
This is important in various physical systems such as
flux lattices in superconductors
\cite{blatter-vortex-review},
charge density waves \cite{gruner-revue-cdw},
colloids and magnetic bubbles
\cite{seshadri-bubbles-thermal}, Wigner crystals
\cite{wigner-millis}, interface roughening
with disorder 
and solid friction of surfaces \cite{hwa-cule}.
An important distinction is whether topological defects
such as dislocations
are present in the system or not. These defects may
appear spontaneously because of disorder or not, or
may be explictly excluded from the model (e.g flux lines
in d=1+1 geometry
\cite{nattermann-flux-creep,carpentier-bglass-layered,nattermann-kierfeld},
tethered networks with permanent
bonds).
Systems which are both elastic and periodic and do {\it not}
contain topological defects
are believed to form glass states with many metastable states,
diverging barriers and slow dynamics
\cite{villain-cosine-realrg,nattermann-pinning,giamarchi-vortex-short,giamarchi-vortex-long,fisher-cdw}.
There are very few analytical methods to study these states
and their physics is far from being completely elucidated,
despite help from extensive numerical simulations.
When topological defects
are present the problem is even more difficult
and very little is known. A crucial question of course
is whether these topological defects, when allowed in
the model, will appear spontaneously at low temperature because of
disorder. 

The general problem of a lattice on a disordered 
substrate was addressed in a previous work
\cite{giamarchi-vortex-short,giamarchi-vortex-long}
and it was predicted that in $d=3$ and for weak pointlike
(uncorrelated) disorder a thermodynamic
phase free of unbound dislocations should exist
and be stable. In this phase, called the Bragg glass, 
translational order decays very slowly beyond a 
translational quasi-order length $R_a$
(at most algebraically). These results 
\cite{giamarchi-vortex-short,giamarchi-vortex-long}
being obtained by mapping the
lattice problem onto a multicomponent version of a random field XY model
the very same predictions (i.e absence of vortices and quasi long range order)
obviously hold for the usual random field XY model as well.
These results are at variance from previous studies
which either argued for \cite{fisher-vortexglass-long}
or assumed \cite{chudnovsky-pinning}
the spontaneous generation of dislocations, or did not address the issue
\cite{blatter-vortex-review,feigelman-collective,fisher-vortexglass-short,nattermann-pinning}.
Predicted consequences \cite{giamarchi-vortex-long}
for the phase diagram of type II superconductors,
i.e existence of a low temperature low field Bragg glass phase
of vortices undergoing a transition into an amorphous
state upon raising the field seem to be in agreement \cite{giamarchi-diagphas}
with the most recent experiments.
The existence of a $d=3$ defect free, Bragg glass type phase 
was confirmed in several numerical simulations either in the
context of superconductors \cite{ryu-diagphas-numerics}
or for the random field XY model \cite{gingras-dislocations-numerics}.
Though a rigorous proof is still lacking
there is also at present further theoretical support for
the stability of the Bragg glass: analytical
variational \cite{carpentier-bglass-layered,nattermann-kierfeld}
and RG studies \cite{carpentier-bglass-layered} in a layered geometry
and, very recently, a proof using improved scaling and energy arguments
\cite{fisher-bragg-proof}.

The Bragg glass is thus an example
of an elastic glass phase with internal periodicity.
In the physics of this type of glasses
two dimensions play a particular role. In addition,
further analytical methods become available
\cite{cardy-desordre-rg,goldschmidt-houghton,toner-log-2}
in $d=2$. When dislocations are {\it excluded} by hand, it is
believed that the lower critical dimension
of these elastic glasses with internal periodicity
is $d_{lc}=2$, with a glass phase existing for $T< T_g$ in $d=2$.
For the $N=1$ component model (random field XY model)
this is the result of the Cardy Ostlund (CO) RG calculation
\cite{cardy-desordre-rg}. By analogy it
can be argued \cite{giamarchi-vortex-long}
that the same holds in the case of the triangular lattice,
but it has not yet been directly verified. A triangular lattice
requires a fully coupled $N=2$ component model. The CO $N=1$ calculation 
could in principle describe a $N=2$ decoupled square lattice,
except that such a lattice has usually a more complicated
elasticity tensor which again couples the components.
When topological defects are allowed, it was 
shown in \cite{cardy-desordre-rg} that for the 
$N=1$ component model they are perturbatively
relevant near the glass transition $T_g$. As argued
in \cite{giamarchi-vortex-long} $T_g$ for the triangular 
lattice is well above the KTNHY melting temperature $T_m$ and 
dislocations should then be relevant near $T_g$
for the $N=2$ triangular lattice as well.
At low temperature however, much less is known 
about the importance of dislocations. The common
belief \cite{blatter-vortex-review},
which is by no means rigorously established, is that
if dislocations are allowed, no glass phase will
exist in $d=2$. This is also hinted at by the
results obtained in \cite{rubinstein-nelson-shraiman}
on the simpler case of the random phase shift model, relevant to
describe a lattice with {\it internal} (i.e structural)
disorder \cite{nelson-elastic-disorder} (a subset of the
present problem). In these early studies
\cite{rubinstein-nelson-shraiman,nelson-elastic-disorder}
the high temperature 
phase with unbound dislocations
was found to be {\it rentrant} at low temperatures
suggesting the importance of topological defects at
low temperature. However, it was pointed out in 
\cite{giamarchi-vortex-long}, from a careful study of the
CO RG flow, that at low temperature the scale 
at which the lattice is effectively dislocation-free
(i.e the distance between unpaired dislocations) can be
{\it much larger} than the translational length $R_a$. Thus
even in $d=2$ the Bragg glass fixed point may be
useful to describe the physics, as a very long crossover
or maybe directly at $T=0$. Furthermore, it was also pointed out
in \cite{giamarchi-vortex-long2} that the conventional
CO RG will not be adequate at low temperature since it 
assumes a thermalized description of the vortices
and neglects important effect such as the
pinning of dislocations by disorder
(i.e that the fugacity depends on
the position).
A similar idea was recently proposed by
Nattermann et al. \cite{natt1} who reconsidered the simpler
random phase shift model. They explicitly showed
that \cite{rubinstein-nelson-shraiman,nelson-elastic-disorder}
was incorrect at low temperature and proposed a
modified coulomb gas RG. It is still an open question
how these new set of ideas and techniques apply to the present
problem.

In the present paper we study the statics and the dynamics
of two dimensional isotropic lattices
interacting with point like disorder {\it excluding
dislocations}. We use a (replica symmetric)
Renormalisation Group (RG) approach as well
as equilibrium dynamics RG.
Though we will not attempt to include
dislocations, it is still interesting for several reasons.
First, it is the natural continuation to $d=2$ of the non trivial
fixed point which describes the Bragg glass phase for $d=3$.
The only other analytical approaches available to describe
the Bragg glass phase are the replica 
variational method
\cite{giamarchi-vortex-short,korshunov-variational-short,giamarchi-vortex-long}
and Fisher's functional RG \cite{fisher-functional-rg}
in a $d=4-\epsilon$ expansion \cite{giamarchi-vortex-short,giamarchi-vortex-long}.
Thus, the study of the present fixed point in $d=2$ is also useful
by comparison.
Second, it is of general interest for systematic study
of the $d=2$ glass phases and allows to show that 
the lower critical dimension is $d=2$.
It is also a first step towards the study of the more difficult problem of
lattices with disorder in $d=2$ with dislocations allowed. Also,
since it has more parameters that the $N=1$ component model
and can be similarly studied numerically, it may lead to
further useful numerical checks of the replica symmetric
RG in this problem as well as the dynamics (see discussion below).
Finally, it may be possible to realize such tethered networks
experimentally. It was argued recently \cite{hwa-cule}
that {\it internal disorder} which breaks the internal periodicity
of the network (and may occur in e.g polymerization) drive the
system to another universality class. Even if this claim,
which we will not investigate here, is correct, crossover lengths 
will probably be large for weak internal disorder.

The $N=1$ Cardy-Ostlund model, which is a subcase of our 
present calculation, has also generated a lot of attention recently,
for different reasons. The gaussian replica variational method (GVM)
applied to this model, leads to a one step replica 
symmetry breaking (RSB) solution below $T_g$ and to mean squared displacements
which grow as $u^2 \sim T_g \ln r$, a different result
from the replica symmetric (RS) RG prediction $u^2 \sim A_1 \ln^2 r$.
Note that the GVM, being by construction an approximation
which neglect some non linearities
has no reason to yield the exact result (see however \cite{korshunov96}).
However it was also shown \cite{ledoussal-rsb-lettre,kierfeld} that the Cardy Ostlund RS-RG flow
is unstable to an infinitesimal RSB perturbation at and below $T_g$.
The issue was thus raised \cite{ledoussal-rsb-lettre} of whether
the RS-RG may miss some of the physics related to RSB. It is thus important to 
perform numerical simulations and carefully check the predictions
of the RS-RG, as well as carefully define the 
observables to be computed. The numerical studies on the 
random phase sine-gordon model presently available show
discrepancies, and their analysis is not satisfactory.
In \cite{numRSGM1} no effect of disorder was found,
while in \cite{numRSGM2} the results seem close to the
predictions of the GVM. In the more recent \cite{numRSGM3,numRSGM4}
the results seem to agree qualitatively with the RS-RG
but with an amplitude {\it much smaller} than the
the prediction for $A_1$ quoted in 
\cite{hwa-fisher-co,batrouni-numerical-cardy}
to which it was compared. Let us also 
point out the two very recent studies \cite{rieger-zerot-co,middleton-zerot-co} of
the random substrate SOS model (believed to be
in the same universality class) directly at
{\it zero temperature} which also yield 
$u^2 \sim A_1 \ln^2 r$. 

In view of the above interest, it is thus important
to compute accurately the amplitude $A_1$ predicted
by the RS-RG. This is one of the result of the present paper.
Our result, smaller by a factor of $4$ from the previous
result \cite{hwa-fisher-co,batrouni-numerical-cardy}
is different from any of the previously published results
\cite{hwa-fisher-co,batrouni-numerical-cardy,goldschmidt-houghton}
and, we argue, is correct. This seems to reconcile,
at least in order of magnitude, the result of the
RS-RG with the result of the numerical simulations 
of \cite{numRSGM3,numRSGM4}. A 
finer and more precise analysis than the one 
performed in \cite{numRSGM3,numRSGM4} is however needed,
including predicted finite size scaling corrections, since
a more careful treatment of the effects of RSB may
reveal that deviations from the RS-RG result are small
\cite{footnote5}
e.g only in the amplitude of the $\log^2 r$ \cite{ledou-giamarchi}.
Another consequence of our results is that the distribution
of rescaled displacements $u/\ln r$ should be {\it gaussian} at large
scale, which we propose as a useful numerical check of the
RS-RG method.

Along these lines, one notes that the
equilibrium dynamical RG studied here,
because it obeys by construction the
Fluctuation Dissipation Theorem (FDT), gives,
as we illustrate explicitly, identical answers for static
quantities than the replica symmetric static RG.
(it thus provides for us a useful check on our statics 
calculations). As discussed above, it is possible
that the static RG does need a RSB structure 
(either strong, i.e in the ground state as in mean field
- or weak, i.e only in the low lying excitations) and a proposal for its
construction was made in
\cite{ledoussal-rsb-lettre}. Also, we interpret the recent results \cite{natt1}
on
the simpler random phase shift model, i.e its mapping 
on some version of Derrida's random energy model,
as another hint that RSB may be important in these
models. If this is the case,
it is then obvious that an equivalent 
statement should exist in the dynamics, i.e that
one may need to construct a RG which violates FDT and
replaces it with a generalized FDT
structure, a program which was successfully carried
in mean field dynamics \cite{meanfield-dynamics1,meanfield-dynamics2}
We will not address this question further
here but will lucidly concentrate on the predictions of
the replica symmetric RG which is certainly a good starting
point in this problem.

The paper is organized as follows. In Section 
(\ref{themodel}) we define the model for a
triangular lattice on a disordered substrate.
In Section (\ref{section3}) 
we study the statics of the problem. First we use
a mapping onto the
replicated Coulomb gas and derive the RG equations
(Section (\ref{section3a}). Then we
analyze the RG flow and compute the
correlation functions (\ref{section3b}). The results
for the triangular lattice can be found in
(\ref{section3b1}). In (\ref{section3b2}) we give the
results for the $N=1$ Cardy Ostlund model and review
the discrepancy with other published results.
In Section (\ref{section4}) we study the dynamics.
In (\ref{section4a}) we carry first order perturbation 
theory and in (\ref{section4b}) we obtain the results
for the dynamical RG. We then compute the dynamical
exponent, and obtain the scaling behaviour of
transport quantities at the transition. In particular we
identify an ``effective critical force'' at the
transition. Section (\ref{section5}) gives the conclusion.
Technicalities are releguated in the appendices. 
In Appendix (\ref{appendixa}) we discuss in detail the regularization,
obtain the RG equations and show explicitely the
cutoff independence of some universal ratios.
In Appendix (\ref{appendixb}) we perform the RG directly on the
static replica effective action. In Appendix (\ref{appendixc}) 
we perform the dynamical RG to second order.
In Appendix (\ref{appendixe}) we study some consequences of the
statistical tilt symmetry.

\section{The model} \label{themodel}

We consider a set of identical ``atoms''. The interactions
between them is such that the ground state configuration in the
absence of disorder is a perfect triangular lattice (see fig. 1)
of equilibrium positions $R_i^0$ and lattice spacing $a_0$.
Both thermal fluctuations and
disorder will result in displacements of the atoms from their
ideal positions $R_i^0$ to new positions $R_i=R_i^0 + u(R_i^0)$.
Furthermore the connectivity of the lattice is fixed: each site
will have exactly six neighbors (see fig 1). This can be realized
in principle by considering a network of identical and permanent 
nearest neighbor bonds (tethers) with the topology of a perfect
triangular lattice. Dislocations are thus excluded
in this model by construction. As discussed in the Introduction,
the model can also be relevant at length scales and in regions of the phase diagram
where dislocations can be neglected.

\begin{figure}
\centerline{\fig{6cm}{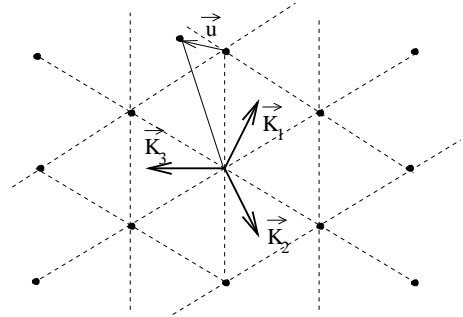}}
\caption{
\label{fig-lattice}
\narrowtext 
Representation of the elastic lattice
and its reciprocal lattice vectors of minimal modulus.}
\end{figure}

The interaction energy can be described in terms of the 
local displacement field $\vec{u}(\vec{r})$ by the harmonic hamiltonian 
\begin{equation}
\label{elastic-interaction}
H_0  =  {1 \over 2} \int {d^2 \vec{q} \over (2 \pi)^2 }
u_i(\vec{q}) \Phi_{ij}(\vec{q}) u_j(-\vec{q}) 
\end{equation}

where $\Phi_{ij}(\vec{q})$ is the elastic matrix of the 2D lattice.
In this paper we will consider isotropic ({\it i.e} triangular) 
lattices which can be described by only two independant elastic coefficients : 
\begin{equation} \label{phi}
\Phi_{ij} (\vec{q})= c_{11} q^2 P_{ij}^L + c_{66} q^2 P_{ij}^T
\end{equation}
 where $P_{ij}^L = \hat{q}_i \hat{q}_j = q_i q_j / q^2$
and $P_{ij}^T= \delta_{ij} - \hat{q}_i \hat{q}_j = (\hat{q}_{\perp})_i(\hat{q}_{\perp})_j$ 
are respectively the longitudinal and transverse projectors.

The interaction of the lattice with the impurity disorder
of the substrate is modelled by adding to the hamiltonian:
\begin{equation}
H_V = \int d^2 \vec{x} ~ \rho(\vec{x}) V(\vec{x}) 
\end{equation}
 where $\rho(\vec{x})$ is the local density of atoms, 
$\rho(\vec{x}) = \sum_i \delta^{(2)}(\vec{x} - \vec{R}_i)$.
The random potential is gaussian with short range correlator
$\overline{V(x) V(x')} = h(x-x')$. The symbol $\overline{..}$ denotes
average over disorder, while $\langle .. \rangle$ denotes
thermal averages.

This model can be transformed, as described in details
in \cite{giamarchi-vortex-long} (see Section II-B), into the
following random phase model:

\begin{eqnarray}
\overline{Z} & = & \int d[u] exp ~ - {1 \over T} \left( H_0 + H_{dis} \right) \\ \label{def-model}
{1 \over T} H_{dis} & = & 
\int_{\vec{x}} \frac{1}{2} \sigma_{ij}(x) (\partial_i u_j + \partial_j u_i) \\ \nonumber
& & + 2 \sqrt{g}  \sum_{\nu} \cos\left( \vec{K}_{\nu}. \vec{u}(x)+\phi_{\nu}(x) \right)
\end{eqnarray}

The first terms corresponds to random local stresses and
comes from the $q \sim 0$ part of the disorder\cite{giamarchi-vortex-long}.
The effect
of only this term
was studied in \cite{nelson-elastic-disorder} (in the presence of dislocations),
and would also arise in a problem of structural
disorder (with no substrate). The second term
originates from the Fourier components of the substrate
potential $V_{K}$ close to one of the reciprocal lattice vectors
\cite{giamarchi-vortex-long}. 
The random field is a phase distributed uniformly
over $[0,2 \pi]$ and satisfies: 
\begin{equation}
\overline{\langle  e^{i(\phi_{\nu}(x)-\phi_{\nu'}(x') } \rangle   }
=\delta_{\nu,\nu'}
\delta^{(2)}(x-x') 
\end{equation}

Note that we are keeping only the terms which are relevant
in the RG sense near $T_g$. There are additional higher
harmonics terms which are irrelevant in $d=2$ near $T_g$. There are also
higher non linear terms which are small in the elastic limit
$|\vec{u}(\vec{R}_{i+1})-\vec{u}(\vec{R}_{i})|\ll a$ where the
derivation of (\ref{elastic-interaction}) is valid and also irrelevant by power counting.
These term will correct the bare values of the relevant 
coupling constants \cite{footnote3}

The random stress field has the general 
correlator:

\begin{eqnarray}
\overline{ \sigma_{ij}(x) \sigma_{kl}(x') } & = & \frac{1}{T}
[ (\Delta_{11} - 2 \Delta_{66}) \delta_{ij} \delta_{kl} \\ \nonumber
&& + \Delta_{66} ( \delta_{ik} \delta_{jl} +
\delta_{il} \delta_{jk} ) ]
\end{eqnarray}
The bare value is $\Delta_{66}=0$ and $\Delta_{11}=\rho_0^2 h_{k=0}/T $
(from  \cite{giamarchi-vortex-long} ) but they will flow under
renormalization.

We will study both the statics and the dynamics of
this model. The statics of this model can be studied using replicas
to average over the disorder potential $V(\vec{x})$
\cite{edwards-replica}. The replicated model reads: 

\begin{eqnarray}
\label{replica-model}
\overline{Z^n} & = & \int d[u^a] exp ~- {1 \over T} \left( H_0^{(n)} + H_I\right) \\
H_0^{(n)} & = & \frac{1}{2} \sum_{ab} \int_q
[ ( c_{11} P_{ij}^L + c_{66} P_{ij}^T ) \delta_{ab} \\ \nonumber
& & - ( \Delta_{11} P_{ij}^L + \Delta_{66} P_{ij}^T ) ] q^2 u^a_i(q) u^b_j(-q) \\
\frac{H_I}{T} & = &  - g \sum_{\nu = 1,2,3} \sum_{a,b=1}^n \cos \vec{K}_{\nu}.
\left[ \vec{u}^a (\vec{x})-\vec{u}^b (\vec{x})\right] 
\end{eqnarray}

$g$ is related to the fourier coefficient
$\Delta_{K_0}$ and the constant density $\rho_0$ :
$g = \rho_0^2 h_{K_0} / T^2 $. We will also use the
dimensionless coupling constant $\tilde{g} = g a^2$.

\section{Study of the Statics using a Coulomb gas formulation} \label{section3}

In this Section we study the statics of the model. We
first derive the renormalization group equations from
a Coulomb gas formulation. In the second part we
compute the static displacement correlation functions.

\subsection{derivation of the RG equations from the Coulomb gas} \label{section3a}

The model (\ref{replica-model}) is the
 analogue for a two-component field of the 
two dimensional random field XY model whose RG equations have been derived using the Coulomb Gas 
approach by Cardy and Ostlund\cite{cardy-desordre-rg}. But contrarily to them, our Coulomb gas 
formulation\cite{nienhuis,nelson-halperin,young} 
is obtained by first using a Villain approximation\cite{villain} before averaging over the disorder.
This results in a natural coupling between the continuous replicated displacement 
field $u^a(\vec{x})$ and vector charges 
$\vec{n}^a_{\nu}(\vec{x})= n^a(\vec{x},\nu).\vec{K}_{\nu}$
 where $\vec{K}_{\nu}$ is one of the three ($ \nu=1,2,3$) reciprocal lattice vector of minimal
modulus (see fig 1) and $ n^a(\vec{x},\nu) \in {\Bbb Z}^n$ :
\begin{eqnarray}
\overline{{Z}^{n}}  =   \int d[u^{a}] \ 
\sum_{[n_{\nu}^{a}(\vec{x})] } \ 
\exp \ \Biggl( && -{1 \over T} H_0^n  -i \ 
\int_{\vec{r}} \vec{n}_{\nu}^{a}.\vec{u}^{a}(\vec{r}) \\ \nonumber
& & +\int_{\vec{r}} \ln (\sqrt{g}) \sum_{a,\nu} 
\left( \vec{n}^{a} \right)^{2} (\vec{r},\nu)\Biggr) 
\end{eqnarray}
with the condition $ \forall (\vec{x},\nu) \ \ \sum_{a} n^{a}(\vec{x},\nu)=0 $. 
The RG equations will be derived to lowest order in the charge fugacity $\sqrt{g}$, since higher order operators 
are irrelevant near the critical point\cite{amit} : thus we can only consider charges of the form 
$n^{a}(x,\nu) =\delta_{\alpha}^{a}-\delta_{\beta}^{a}\equiv \delta_{\alpha,\beta}^a$ where $\alpha$ and $\beta$ 
are replica integers between $0$ and $n$. 
 The effective fugacity of these minimal charges is $g$. 
Integrating over the smooth field $u(\vec{x})$ one recover a 2D vector Coulomb Gas
 with charges having both a spatial and a replica structure, and whose action 
is (omitting the fugacity term):

\begin{eqnarray}
S[n] & = & { T \over 2}\sum_{\nu,\nu'} \int_{\vec{q}}  \vec{n}^{a}_{\nu,i}(q)
(\Phi^{-1})_{ij}^{ab} \vec{n}^{b}_{\nu',j}(-q) \\
& = & - { T \over 2}\sum_{\nu,\nu'} \int_{\vec{r},\vec{r}'}
\vec{n}_{\nu,i}^{a}(\vec{r})V_{ij}^{ab}(\vec{r}-\vec{r}')\vec{n}_{\nu,j}^{b}(\vec{r}')
\end{eqnarray}

 where the interaction $V(\vec{r})$  is obtained by a 
 Fourier transform. We have incorporated the replica
off diagonal terms in $\Phi^{-1}_{ab}$.

 In this section, 
as is usualy done, we will use\cite{footnote6}, instead of 
the full interaction, its asymptotic form\cite{nelson-halperin}
 ($r\gg a$) (see Appendix A) :

\begin{equation}
\label{CG-interaction}
V_{ij}(\vec{r}) = \delta_{ij} \left( {\kappa}^{ab}_{1} \ln {r \over a} + \frac{1}{2} {\kappa}^{ab}_{2}\right)
\ \ -{\kappa}^{ab}_{2}\ { r_{i}r_{j} \over r^{2} }
\end{equation}
and ${\kappa}^{ab}_{1} =   \left( c_{66}^{-1} +c_{11}^{-1}\right)^{ab}/4 \pi$, 
${\kappa}^{ab}_{2} =  \left( c_{66}^{-1} -c_{11}^{-1}\right)^{ab}/4 \pi $. We will use the notation 
${\kappa}^{ab}_{1}= {\kappa}_1 \delta_{ab} -\overline{{\kappa}_1}$ and 
$\left( c_{11}\right)^{ab}= c_{11}\delta^{ab} -\Delta_{11}$. 

We won't deal with possible appearence of dislocations in this model.
This will lead to an electromagnetic vector CG 
which neads more work\cite{work-in-progress}. This simplification allows us to derive RG equations by working 
on the 2-point correlation function\cite{nelson-in-domb} instead of the replicated partition 
function\cite{young,nelson-halperin}. These two 
schemes are equivalent\cite{footnote} to order (at least) $g^2$. 
The renormalised elastic matrix can be defined as 
\begin{eqnarray}
 q^2 \left(\Phi_{R}^{-1}\right)_{ij}^{ab} & = &  (c_{11,R}^{-1})^{ab} P_{ij}^{L} +
(c_{66,R}^{-1})^{ab} P_{ij}^{T}  \\
&= & \lim_{q \rightarrow 0} { q^{2}\over T} \langle  u_{i}^{a}(\vec{q})u_{j}^{b}(-\vec{q})\rangle   
\end{eqnarray}
 To be consistent this definition must be independant of the direction $\hat{q}$ : this is simply a direct consequence 
of the isotropic nature of the 2nd order invariant tensors on the triangular lattice. 
The explicit expression for the renormalised elastic coefficients naturally follows :
\begin{eqnarray}
\label{renorm-elastic}
& &  \left(c_{11,R}^{-1} -c_{11}^{-1}\right)^{ab} =  \\ \nonumber 
& & -T \lim_{q\rightarrow 0} q^{2}
\langle   \vec{n}^{c}_{\nu,l}(\vec{q}) \vec{n}^{d}_{\nu',m}(-\vec{q})\rangle  _{S[n]}
P_{ij}^{L}
\left(\Phi^{-1}\right)_{il}^{ac}\left(\Phi^{-1}\right)_{jm}^{bd}
\end{eqnarray}
 where $i,j,l,m$ are cartesian coordinates, and we used the notation $ \vec{n}^{c}_{\nu,l} = n^c \times \left(\vec{K}_{\nu}\right)_l$.   
 The same holds for $c_{66}^{-1}$ by simply replacing $P_{ij}^L$ by $P_{ij}^T$. 

\begin{figure}[htb]
\centerline{ \fig{6cm}{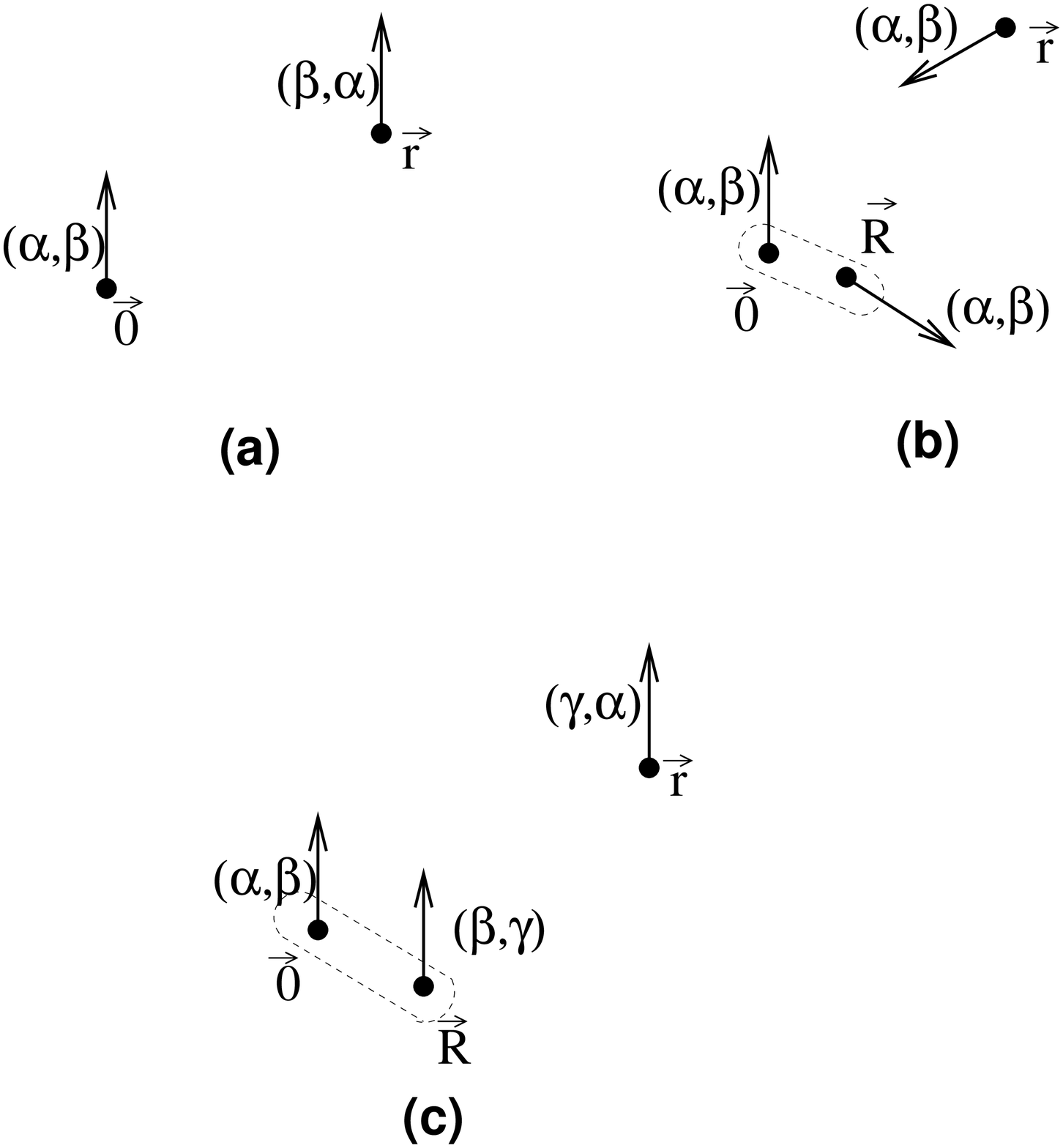} }
\caption{
\label{fig-correl}
\narrowtext 
 Representation of the three different configurations that contribute 
to the correlation function of the Coulomb Gas charge $\vec{n}$. Configuration (a) is of 
order $g^{2}$ whereas (b) and (c) are of order $g^{3}$.
 }
\end{figure}

Replacing $\Phi^{-1}$ by its expression (\ref{phi}) and using the Fourier transform of the charge correlation, we obtain 
\begin{eqnarray}
\label{perturb-elastic}
\left(c_{11,R}^{-1}\right)^{ab} = 
\left(c_{11}^{-1}\right)^{ab} & + & 
{T \over 2} P_{ij}^L \left( c_{11}^{-1}\right)^{ac}\left( c_{11}^{-1}\right)^{bd} \\ \nonumber
& \times & \int_{\vec{r}} (\hat{q}.\vec{r})^{2} 
\langle  \vec{n}_{\nu,i}^{c}(\vec{0})\vec{n}_{\nu',j}^{d}(\vec{r})\rangle_{S[n]}
\end{eqnarray}

This equation is the starting point of our RG study. Note that itis a definition of the renormalised 
elastic coefficient exact to all order in $g$. We now proceed as follow : first we expand perturbatively in $g$ 
the right-hand side of this 
equation. One can then apply 
a coarse-graining procedure which leaves 
the form of this perturbative expansion unchanged, thus defining coarse-grained elastic coefficients. 
To order $g^3$ only three terms appear (see fig.2) : (a) of order $g^2$ and (b) and (c). 
 Corrections coming from the coarse-graining of 
configuration (a) defines the coarse-grained elastic constant 
 $c_{11}^{-1}(l)$ to 
order $g^2$ and $g(l)$ to order $g$, where $l$ is the 
scaling parameter. In the spirit of Coulomb Gas 
renormalisation\cite{nienhuis}, this corresponds to the annihilation of charges 
(screening of the coulomb interaction), whereas the 
fusion of charges is given by the contributions of configurations (b)
and (c) when $|\vec{R}|\ll |\vec{r}|$ or when $|\vec{R}-\vec{r}|\ll |\vec{r}|$. This gives 
the correction of 
$g(l)$ to order $g^2$. 

 First, let us develop the two point correlation function 
$\langle  \vec{n}_{\nu,i}^{c}(\vec{0})\vec{n}_{\nu',j}^{d}(\vec{r})\rangle$ to order $g^3$ : 
it is given by the three contributions (a), (b), 
and (c) (see fig 2) : 
\begin{equation}
\label{correl-real}
\langle  \vec{n}_{\nu,l}^{c}(\vec{0})\vec{n}_{\nu',m}^{d}(\vec{r})\rangle_{S[n]} = 
g^2 {\cal A}_1 (\vec{0},\vec{r})+ g^3 {\cal A}_2 (\vec{0},\vec{r})
\end{equation}
 where the first term corresponds to the only neutral 2-charges configuration : 
\begin{equation}
{\cal A}_1 (\vec{0},\vec{r})  =  \sum_{\nu}\sum_{\alpha\neq\beta}^n \delta^c_{\alpha,\beta}
\delta^d_{\beta,\alpha}(K_{\nu})_l(K_{\nu})_m e^{-S_a} 
\end{equation}
 and configurations (b) and (c) contribute to the factor ${\cal A}_2$ : 
\begin{eqnarray*} 
{\cal A}_2 (\vec{0},\vec{r}) & = & 
2 \int_{\vec{R}} \sum_{\nu\neq \nu'}\sum_{\alpha\neq\beta}^n 
\delta^c_{\alpha,\beta} \delta^d_{\alpha,\beta} (K_{\nu})_l(K_{\nu'})_m e^{-S_b} \\
 & + &  
2 \int_{\vec{R}} \sum_{\nu}\sum_{\alpha\neq\beta\neq \gamma }^n 
\delta^c_{\alpha,\beta} (\delta^d_{\beta,\gamma}+\delta^d_{\gamma,\alpha})   
(K_{\nu})_l(K_{\nu})_m e^{-S_c}
\end{eqnarray*}

 In these expressions, $S_a,S_b,S_c$ are the respective actions of the three
 configurations : 
$ S_a = 2T |\vec{K}|^2 \left( {\kappa}_1 \ln (r / a)
- {\kappa}_2 (\hat{K}.\hat{r})^2 +\kappa_2/2 \right) $.
 The two others can be easily expressed in terms of $S_a$. 
We can put this back in the expression (\ref{perturb-elastic}) of $c_{11,R}^{-1}$. 
The angular integral of the 
$g^2$ term can be performed, using the parameter $\alpha = 
{\kappa}_2 |\vec{K}|^2 T$ :
\begin{eqnarray}
\label{def-alpha}
B_{\stackrel{{\scriptstyle 11}}{\scriptstyle 66}}(\alpha) 
& = &\sum_{\nu}  \int d\hat{r} 
 \left( \hat{q} . \hat{r} \right)^2  
\left( \hat{q}_{\stackrel{{\scriptstyle \parallel}}{\scriptstyle \perp}} . K_{\nu} \right)^2 
e^{2 \alpha  (\hat{K}.\hat{r})^2 -\alpha} \\
& = &  K^2 {3 \pi \over 4} \left(\pm  I_1(\alpha) + 2 I_0(\alpha) \right) 
\end{eqnarray}

 where $I_0$ and $I_1$ are the modified Bessel functions. 
The function $B_{11}$ appears in the expression of $c_{11,R}^{-1}$ 
while $c_{66,R}^{-1}$ involves $B_{66} $.
 Using the following results on the replicated charges 
\begin{eqnarray*}
&&\sum_{\alpha,\beta} \delta^a_{\alpha,\beta}\delta^b_{\beta,\alpha}=
-2\left(n \delta^{ab}-1\right)\\
&&\sum_{\alpha\neq\beta\neq \gamma}\delta^a_{\alpha,\beta}\delta^b_{\beta,\gamma}= (2-n)\left(n \delta^{ab}-1\right)
\end{eqnarray*}

 and the definition $\tilde{g}=g a^2$, we can express $c_{11,R}^{-1}$ as 

\begin{eqnarray}
\label{develop}
&& \left( c_{11,R}^{-1} \right)^{ab} =  \left( c_{11}^{-1} \right)^{ab}(l) \\ \nonumber
&-& { T \over c_{11}^2} \tilde{g}^2(l)
(n\delta^{ab}-1)B_{11}(\alpha) \int_{a}^{\infty} {dr \over a} 
\left( {r \over a} \right)^{3-2 |\vec{K}|^2 K_1 T} \\ \nonumber
\nonumber
 & & + {T \over c_{11}^2} \tilde{g}^3(l)(n\delta^{ab}-1) 
\int_{|\vec{r}|\geq a} {d^2 \vec{r} \over a^2 } K^2 \Lambda(\vec{r})
\end{eqnarray}

  The integrand in the last term is given by :

\begin{eqnarray*}
\Lambda(\vec{r}) = \int_{|\vec{R}|\geq a} {d^2 \vec{R} \over a^2 } {3 \over 8} && \left({r\over a}\right)^2
\left( \cos 2(\hat{K},\hat{r}) + 2 \right) \\ \nonumber
 & & \times \left(  (2-n) e^{-S_c}- e^{-S_b}\right)
\end{eqnarray*}

 We can then apply the coarse-graining procedure to this self-consistent equation. We thus rescale 
the hard-core cut-off from $a$ to $\tilde{a}=ae^{dl} \approx a + adl$. The first integral split into two terms 
$(\eta =|\vec{K}|^2 {\kappa}_{1} T )$ : 
\begin{eqnarray*}
 \int_{a}^{\infty}{dr \over a} \left( {r\over a}\right)^{3-2\eta} 
 & \longrightarrow  & \int_{a}^{ae^{dl}} \frac{dr}{a} \left( {r\over a}\right)^{3-2\eta}
 +  \int_{ae^{dl}}^{\infty} \frac{dr}{a} \left( {r\over a}\right)^{3-2\eta} \\
& \approx & dl + e^{l(4-2\eta)} 
 \int_{\tilde{a}}^{\infty} \frac{dr}{\tilde{a}} \left( {r\over \tilde{a}}\right)^{3-2\eta}
\end{eqnarray*}

 Hence the first terms contribute to $c_{11}^{-1}(l)$ while the second renormalises
$\tilde{g}(l)$ :
\begin{eqnarray}
\label{RG-eqs}
d\ [c_{11}^{-1}]^{ab} (l) & = &  - \tilde{g}^{2} (n \delta^{ab}-1)
 T B_{11}(\alpha)c_{11}^{-2}~dl\\
d \ \tilde{g}^2(l) & = &  (4- 2 |\vec{K}|^2 {\kappa}_{1} T) \tilde{g}^2 ~dl
\end{eqnarray}

  For the purpose of this paper, we are only interested in the contribution of the 
last term of eq. (\ref{develop}) to $dg^2$. It corresponds to configuration (b) and (c) of 
(fig.2), when $a \leq |\vec{R}|\leq a(1+dl)$ or $a \leq |\vec{R}-\vec{r}|\leq a(1+dl)$. 

The renormalisability of the model that one can directly see 
in our procedure ensures that this correction can be written as a a contribution 
to $\tilde{g}^2(l)$ of order 
$\tilde{g}^3$. First of all we check these renormalisation properties of the hamiltonian : 
$$ \lim_{|\vec{R}| \rightarrow a} e^{-S_b(\vec{r},\vec{R}) }=
e^{- S_a (\vec{r})} \ 
e^{-2 \alpha (\hat{K}_1.\hat{R})(\hat{K}_2.\hat{R})  -\frac{\alpha}{2}}$$

$$\lim_{|\vec{R}| \rightarrow a} e^{-S_c(\vec{r},\vec{R})}=
e^{-S_a(\vec{r})} \ 
e^{\alpha (\hat{K}_1.\hat{K})^2-\frac{\alpha}{2}}$$

 The corresponding correction is 
\begin{equation}
\delta \Lambda(\vec{r}) = 
 {3 \over 4} 
\left( {r \over a}\right)^2 \left( \cos 2(\hat{K},\hat{r}) + 2 \right) e^{-S_a} 
B_g(\alpha)
\end{equation} 

 where the fonction $B_g(\alpha)$ is given by 

\begin{eqnarray}
B_g(\alpha)  &  = & 
\int d\hat{R} e^{-\alpha/2}
[ (2-n) e^{\alpha \cos^2 (\hat{K},\hat{R})} \\ \nonumber
& & ~~~~~~~~~~~~~~~- e^{-2 \alpha \cos(\hat{K},\hat{R})
\cos(\hat{K}',\hat{R}) } ] \\ \nonumber
 & = &  
 2 \pi
\left[ (2-n) I_0(\alpha /2) - I_0(\alpha) \right]  
\end{eqnarray}
Integration over $\vec{r}$ leads to
$$
  \int {d^2 \vec{r} \over a^2} K^2 \delta \Lambda(\vec{r})  =  
2 B_g(\alpha)B_{11}(\alpha) \int_{a}^{\infty}
{dr \over a} \left( {r \over a} \right)^{3-2K^2 {\kappa}_{1}  T}
$$

 This is exactly the expected form, which can be written as a correction to $g^2$ :

\begin{equation} 
d(g^2)= - dl~ \tilde{g}^3 2 B_g(\alpha)
\end{equation}
where 
$$ B_g(\alpha) =- 2 \pi \left ( I_0(\alpha) + ( n-2) I_0(\alpha/2) \right) $$
 The final equations, obtained after the $n\rightarrow 0$ limit, are
\begin{mathletters} \label{rg1}
\begin{eqnarray}
\label{RG-final}
{d\ c_{11} \over d\ l} & = & {d\ c_{66} \over d\ l} =0 \\
{d\ \Delta_{11}(l) \over  d \ l} & = &  \tilde{g}^{2} T B_{11}(\alpha) \\
{d\ \Delta_{66}(l) \over  d \ l} & = &  \tilde{g}^{2} T B_{66}(\alpha) \\
{d \ \tilde{g} \over d\ l} & = &  (2-  K^2 {\kappa}_{1}  T ) \tilde{g} -
\tilde{g}^2  B_g(\alpha)
\end{eqnarray}
\end{mathletters} 
 with $\tilde{g}=g a^2$. From now on 
$B_g(\alpha)$ denotes its value at $n=0$.

It is useful to compare these equations with the
one for the $n=1$ component model of Cardy and Ostlund.
To obtain them we set $c_{11}=c_{66}=c$
(i.e $\alpha=0$),
$\Delta_{11}=\Delta_{66}=\Delta$ and 
consider the two reciprocal lattice vectors 
(instead of three) of a square lattice
in the interaction term with $K^2=1$.
This gives two decoupled $n=1$ component model
with RG equations obtained by $B_{11}= B_{66} \to \pi$ and
drop the term not proportional to $n-2$ in the $g^2$ correction.

\begin{mathletters}
\label{RG-co}
\begin{eqnarray}
{d c \over dl} & = & 0 \\
{d \Delta(l) \over  dl} & = &  \pi \tilde{g}^2  T  \\
{d \tilde{g} \over d\ l} & = &  (2 -   \frac{T}{2 \pi c} ) \tilde{g}  -
4 \pi \tilde{g}^2 
\end{eqnarray}
\end{mathletters}
the transition temperature is thus $T^{CO}_g = 4 \pi c$.

\subsection{Analysis of the RG flow and static correlation functions}
\label{section3b}

We now study the RG flow and compute the correlations.
In the case of the $N=1$ model this was first done in
\cite{cardy-desordre-rg,goldschmidt-houghton} and later
reconsidered in \cite{toner-log-2,denis-bernard}.

\subsubsection{triangular lattice}
\label{section3b1}

The flow defined by the above RG equations (\ref{rg1}) has similarities 
with the one obtained for the random field XY model\cite{cardy-desordre-rg}.
There is a transition at $T_g = 2/( K^2 \kappa_1) = 8 \pi K^{-2} c_{11} c_{66}/
(c_{11} + c_{66} )$ and one has $K^2 = 16 \pi^2/3 a_0^2$ for the
triangular lattice. 

In the high temperature phase $T> T_g$ the disorder renormalizes to zero.
At low temperature $T < T_g$ the coupling constant $\tilde{g}(l)$ converges
towards a perturbative fixed point $\tilde{g}^*$. Introducing the reduced temperature
$\tau = (T_g - T)/T_g $, one has:
$\tilde{g}^* = 2 \tau/B_g(\alpha)$
with $ B_g(\alpha) = 2 \pi ( 2 I_0(\alpha/2) - I_0(\alpha) )$.
It depends continuously on the
value of $\alpha$, defined in (\ref{def-alpha}) and which we can evaluate at
$T_g$ since we work near $T_g$. Using the above value for 
$T_g$ one finds that
$\alpha(T_g) = \alpha_g = 2 (c_{11}-c_{66})/(c_{11}+c_{66})
= 2 (1+\sigma)/(3-\sigma)$ where $\sigma$ is the Poisson ratio
defined as usual as $\sigma = 1 - 2 c_{66}/c_{11} = \lambda/(\lambda + 2 \mu)$.
Thus we find that the obtained fixed point depends continuously
on the Poisson ratio. Since the Poisson ratio is not renormalized
one has now a {\it plane} of fixed points, rather than a line
in the case of the $N=1$ Cardy Ostlund model, parametrized 
by the temperature $\tau$ and the Poisson ratio $\sigma$.

One must check however that the perturbative fixed point does indeed exist,
i.e that $ B_g(\alpha) >0$ on the allowed domain of
variations of $\alpha$. This is indeed the case since one
finds that $ B_g(\alpha) >0$
as long as $\alpha \leq \alpha^*\approx 2.218$. This condition is fullfilled
since $c_{66} >0$ implies that $\alpha <2$.

At this fixed point both $\Delta_{11}(l)$ and $\Delta_{66}(l)$ 
grow to non perturbative values. This is not a problem since due to 
statistical symmetry \cite{statistical-inv} they do not feedback
in the RG equations.
This produces however a change in the
correlation functions (see e.g Appendix \ref{appendixe}).

We can integrate the above RG equations
and the solution reads:

\begin{eqnarray} \label{solution-RG}
 \tilde{g}(l) &=& \frac{ g_0 e^{ 2 \tau l} }{ 1 + \chi (e^{2 \tau l} -1) } \\
 \Delta_{11,66} (l)&=& \Delta(0) +
\frac{D_{11,66}}{2 \tau} \Biggl( \ln ( 1 + \chi (e^{2 \tau l} -1) ) \\ \nonumber
& & -\chi( 1 -\chi )
\frac{ e^{2 \tau l} -1 }{ 1 +  \chi(e^{2 \tau l} -1) } \Biggr)
\end{eqnarray}

where we defined $D_{11,66}= 4\tau^2 T B_{11,66}(\alpha) / B_g^2(\alpha)$ 
and $\chi = g_0 B_g(\alpha)/(2 \tau)$. Note that $\chi =g_0/\tilde{g}^*$
for $\tau>0$. Thus at large
$l$ one has in the glass phase ($\tau > 0$):

\begin{eqnarray}
&& \tilde{g}(l) = \tilde{g}^* + {\cal O}(e^{-2 \tau l}) \\ \label{delta-glass}
&& \Delta_{11,66}(l) =  D_{11,66} \ln \left( \frac{a e^l}{\xi} \right)
+ \Delta_{11,66}(0) + {\cal O}(e^{-2 \tau l}) 
\end{eqnarray}

where we have introduced the length:
\begin{eqnarray}
\xi = a \exp \left(  \frac{1 - \chi - \ln \chi}{2 \tau} \right)
\end{eqnarray}

For weak disorder $g_0 \ll 2 \tau/B_g$ it corresponds to the
length $\xi \sim R_a \sim (2 \tau/g_0)^{1/(2 \tau)} \gg a$
at which translational order decays asymptotically (also equal 
to the Larkin length $R_c$
for this lowest harmonic model \cite{carpentier-bglass-layered},
valid near $T_g$). Note that one can define another
length scale $\xi_c$ defined as the crossover length between a 
$\log(r)$ behaviour of the mean squared displacements (see below) 
for $r < \xi_c$ to a large $r$ regime 
which corresponds to a $\log^2(r)$ asymptotic correlation.
This length diverges exponentially as $\xi_c \sim exp(- cst /\tau^2)$ 
at the transition $\tau=0$.

We now compute the correlation functions \cite{toner-log-2}. We will
be interested in the following correlation functions:

\begin{eqnarray}
B_{ij}(r) & = & 
\overline{ \langle ( u_i(\vec{r})-u_i(\vec{0}) ) ( u_j(\vec{r})-u_j(\vec{0}) ) \rangle } \\ \nonumber
 & = & 
2 \int {d^2 \vec{q} \over (2 \pi )^2 }  
(1-e^{i\vec{q}.\vec{r}}) \overline{ \langle u_i(\vec{q})u_j(-\vec{q}) \rangle } \\ \nonumber
& = & B_L(r) \tilde{P}_{ij}^L +  B_T(r) \tilde{P}_{ij}^T
\end{eqnarray}
 where the projectors are defined by $\tilde{P}_{ij}^L = \hat{r}_i \hat{r}_j$ and 
$\tilde{P}_{ij}^T = (\hat{r}_{\perp})_i (\hat{r}_{\perp})_j$.
 The longitudinal correlator is thus given by 
\begin{equation}
B_L(r) = 2 \int { d^2 \vec{q} \over (2 \pi)^2 }  
\left( 1- e^{i \vec{q}.\vec{r}}\right) \tilde{P}_{ij}^L 
\overline{ \langle u_i(\vec{q})u_j(-\vec{q}) \rangle }
\end{equation}
The transverse correlation function follows by replacing $\tilde{P}^L$ by $\tilde{P}^T$. 

We define 
\begin{equation}
\Gamma_{ij}(\vec{q},\Delta_{11,66}(0),\tilde{g}(0))=
\overline{ \langle \vec{u}_i(\vec{q})\vec{u}_j(-\vec{q}) \rangle }
\end{equation}

 Using usual dimensional scaling relations, we can write 
$$ \Gamma_{ij}(\vec{q},\Delta_{11,66}(0),\tilde{g}(0)) = e^{2l}~\Gamma_{ij}(e^l \vec{q},\Delta_{11,66}(l),\tilde{g}(l))$$

 We then choose the scaling parameter $l$ such that\cite{toner-log-2}
  $qe^l = 1/a$. The large $r$ behaviour of the correlation 
function thus corresponds to the limit $e^l \rightarrow \infty$. The RG flow approaches its fixed point 
in that limit : $\tilde{g} \rightarrow \tilde{g}^{\infty}\approx \tilde{g}^*$.
Assuming that $\Gamma_{ij}(1/a,\Delta_{11,66}(l),\tilde{g}^*)$ can be
evaluated perturbatively in $\tilde{g}^*$ near $T_c$
We can then evaluate the correlation function : 
\begin{eqnarray}
&&\Gamma_{ij}(\vec{q},\Delta_{11,66}(0),\tilde{g}(0)) = \\ \nonumber
&& {T \over q^2} 
 \left( 
(c_{11}^{-1} +{\Delta_{11}(l) \over  c_{11}^{2}} ) P^L_{ij} + (c_{66}^{-1} +{\Delta_{66}(l) \over c_{66}^2}) 
P^T_{ij} \right)
\end{eqnarray}

The correlation function then takes the following form near $T_g$:
\begin{eqnarray}
\overline{ \langle \vec{u}_i(\vec{q})\vec{u}_j(-\vec{q}) \rangle } \stackrel{q \rightarrow 0}{\simeq} & &
 -{T_g \over q^2} \ln ( q  \xi) 
\left( { D_{11} \over c_{11}^{2}}  P_{ij}^L
+  {  D_{66} \over c_{66}^{2}} P_{ij}^T \right)     
\end{eqnarray}

The angular integrals can be easily performed using the formula:
\begin{equation}
\label{angular-int-0}
{1 \over 2 \pi} \int d\hat{q} ( 1-e^{i \vec{q}.\vec{r}}) P_{ij}^{\alpha}
\tilde{P}_{ij}^{\beta} =
{1 \over 2} \left( 1-J_0(qr) + \epsilon_{\alpha \beta} J_2(qr)\right)
\end{equation}

with $\alpha, \beta = L,T$ and $\epsilon_{LL}=\epsilon_{TT}=1$ and
$\epsilon_{LT}=\epsilon_{TL}=-1$. 

This gives:

\begin{eqnarray}
\label{angular-int}
B_{L,T}(r) & = & {1 \over 2 \pi} (\tilde{B}_{11} + \tilde{B}_{66})
\int \frac{dq}{q} \ln(1/(q \xi)) (1-J_0(qr)) \\ \nonumber
& &\pm 
{1 \over 2 \pi} (\tilde{B}_{11} - \tilde{B}_{66})
\int \frac{dq}{q} \ln(1/(q \xi)) J_2(qr)
\end{eqnarray}
 where $\tilde{B}_{11,66}= T_g D_{11,66}/c_{11,66}^2$. 
The integrals can be evaluated as:

\begin{eqnarray*}
& & {1 \over 2 \pi} 
\int_0^{{2 \pi\over a}} dq {\ln (q \xi) \over q} \left( 1-J_0(qr)\right) \\ 
& &= -{1 \over 4 \pi} \ln^2\left( {r \over \xi} \right)
+ {\cal O}\left( \ln \left( {r \over a} \right) \right) \\
& & {1 \over 2 \pi} 
\int_0^{{2 \pi\over a}} dq {\ln (q \xi) \over q} J_2(qr) \\
&& = -\frac{1}{4\pi} \left( \ln \left( {r \over \xi}\right) \right) 
 + {\cal O}\left( 1 \right)
\end{eqnarray*}

The difference between the expressions of $B_L$ and $B_T$ is the sign of the second order 
Bessel function $J_2(qr)$. 
Both these correlations have the same leading term in the limit of large $r$, given 
to the lowest order in $\tau$ by:

\begin{eqnarray}
\label{correl-long}
B_L(r) \sim B_T(r) \sim \frac{b(\alpha)}{K^2} \tau^2  \ln^2\left( {r \over \xi} \right)
\end{eqnarray}

with a universal coefficient (the cut-off dependence drops out)
$b(\alpha) = (\tilde{B}_{11} + \tilde{B}_{66})/(4 \pi)$
Another universal quantity (in the limit of large $r$)
is the difference $B_T(r)-B_L(r)$ since
the leading integral, with all its cutoff dependent
corrections in $\ln r/a$ cancel exactly. One obtains:

\begin{equation}
\label{correl-trans}
B_T(r)-B_L(r) \sim \frac{\tilde{b}(\alpha)}{K^2}  \tau^2
\ln \left( {r \over \xi}\right) 
\end{equation}

The coefficient of the log is a universal function of $c_{11}$ and $c_{66}$, 
$\tilde{b}(\alpha) = (\tilde{B}_{66} - \tilde{B}_{11})/(4 \pi)$.
Specifically one finds (see fig.3 for a plot of these functions) :

\begin{eqnarray}
&& b(\alpha) = 6
\frac{ 2 I_0(\alpha) (1 + \frac{\alpha^2}{4}) - \alpha I_1(\alpha) }{
(2 I_0(\alpha/2) - I_0(\alpha) )^2 } \\
&& \tilde{b}(\alpha) = 6
\frac{ I_1(\alpha) (1 + \frac{\alpha^2}{4}) - 2 \alpha I_0(\alpha) }{
(2 I_0(\alpha/2) - I_0(\alpha) )^2 } \\
\end{eqnarray}

where $\alpha = \alpha_g = 2 (c_{11}-c_{66})/(c_{11}+c_{66})$.

Here it is interesting to remark that in addition to the two terms (\ref{correl-long},\ref{correl-trans}), 
$B_{L,T}(r)$ contains a non universal term proportional to $\ln (r/a)$
(e.g a term like $\Delta(0) \ln (r/a)$).
Therefore when comparing 
the RG predictions with e.g numerical results, one should be careful in identifying 
the different components of the spatial correlations.

\begin{figure}
\centerline{\fig{8cm}{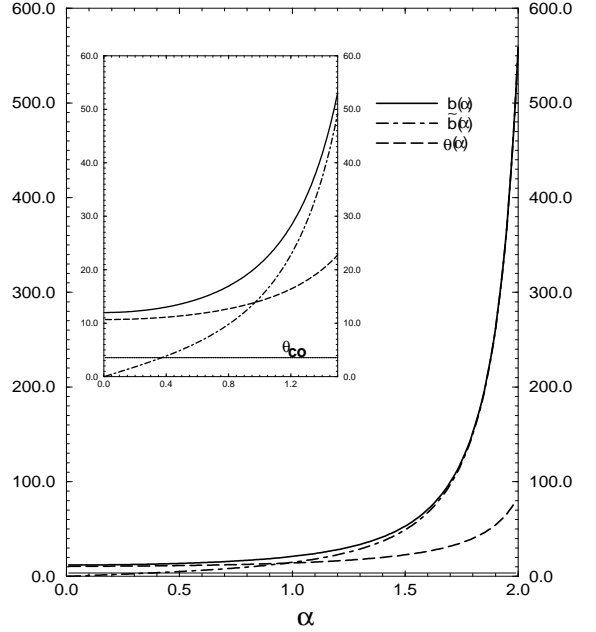}}
\caption{
\label{fig-RGcoef}
\narrowtext 
Plot of the numerical coefficients of the correlation functions $B_{T,L}(r)$ 
 and $B_T(r)-B_L(r)$  : $b(\alpha)$ and $\tilde{b}(\alpha)$ as function of the elastic ratio 
$\alpha = 2(1+\sigma)/(3-\sigma)$. We have also plotted the coefficient of the dynamical exponent :
 $\theta(\alpha)= (z-2)/ \tau$. $\theta_{co}$, the corresponding value for the planar random field 
XY model (the Cardy-Ostlund model) is shown as a reference. See 
the text for the details.}
\end{figure}

As we discuss in the Appendix \ref{appendixe}
the distribution of rescaled 
displacements $u/\ln r$ should be {\it gaussian}. Thus we have also obtained
the decay of the structure factor at large $r$:

\begin{eqnarray}
C_{K_0}(r) = \overline{ e^{i K_0.u(r)} e^{- i K_0.u(0)} } \sim
\exp( - \frac{1}{2} b(\alpha) \tau^2  \ln^2( {r \over \xi}) )
\end{eqnarray}

the corresponding result for the $N=1$ CO model
structure factor is given in \ref{result-structure} 
of the Appendix \ref{appendixe}.
Though the $\ln^2 r$ term is isotropic there is 
subdominant anisotropy. For instance one has
for the decay in different directions:

\begin{eqnarray}
\frac{C_{K_0}(r \parallel K_0)}{C_{K_0}(r \perp K_0)} 
\sim (\frac{\xi}{r})^{\frac{\tilde{b}(\alpha) \tau^2}{2}} 
\end{eqnarray}

which is the analogous for $d=2$ of Eq. (4.32) of Ref.
\cite{giamarchi-vortex-long}.

{\it High temperature phase}

Finally, the high temperature phase is characterised 
by $\tilde{g}(l\rightarrow \infty)=0$ and a non universal 
value of $\Delta_i(l)\rightarrow \Delta^{\infty}_i$. From (\ref{solution-RG})
we find 
\begin{equation}
\Delta^{\infty}_{11,66} = \Delta_{11,66}(0) + \frac{D_{11,66}}{2\tau} \left( \ln(1-\chi) +\chi\right)
\end{equation}
 which depends on the initial bare values of $\Delta$
but remains finite at the transition $\tau \to 0^{-}$.

Since the renormalised theory is gaussian, the correlation 
 function is straightforwardly given in the limit of large $r$
by:
\begin{eqnarray}
 && \overline{ \langle ( u_i(\vec{r})-u_i(\vec{0}) ) ( u_j(\vec{r})-u_j(\vec{0}) ) \rangle }   \sim  \\ \nonumber
 && {T \over \pi} \delta_{ij} \ln \left( { r \over a}\right)
  \left( 
 (c_{11}^{-1} +{\Delta_{11}^{\infty} \over  c_{11}^{2}} )  + (c_{66}^{-1} +{\Delta_{66}^{\infty} \over c_{66}^2}) 
  \right)
 \end{eqnarray}

\subsubsection{Cardy-Ostlund model}
\label{section3b2}

In the case of the Cardy Ostlund $N=1$ problem, defined in its
replicated version, as:

\begin{eqnarray}
\label{cardyostlund}
\frac{H}{T} & = & \sum_{ab} \int_r ( \frac{1}{2 T}
(c \delta_{ab} - \Delta_{ab}) \nabla u_a \nabla u_b
- g \cos(K (u_a - u_b))
\end{eqnarray}

one finds, using (\ref{RG-co}):

\begin{eqnarray}  \label{result-co}
B(r) = \overline{\langle (u(r) - u(0))^2 \rangle } 
\sim \frac{2}{K^2} \tau^2 \ln^2 (r/\xi)
\end{eqnarray}
where, we recall $\tau = (T_g-T)/T_g$. 
As announced in the Introduction, this result is
different from all the results 
previously published (to our knowledge). Since it is important for
comparison with several existing numerical simulations,
we now discuss in more details this discrepancy.

Our result is smaller by a factor of $2$ from the
original result (5.24) of Ref. \cite{goldschmidt-houghton}.
It is smaller by a factor of $4$ from the result
quoted in Ref. \cite{hwa-fisher-co}, and still by a factor
of $2$ from its later corrected value \cite{hwa-privcomm}
(it is also larger by a factor of $4$ from the result obtained
in \cite{denis-bernard}).

As discussed above the RG equations which read at
$T_g$, $d\Delta/dl = c_1 g^2$, $dg/dl = c_2 g^2$
are not universal
but the following amplitude ratio, defined in 
\cite{hwa-fisher-co} 
$R = T_g c_1/(c c_2)^2$ is universal.
We find here that $R=\pi$. 
(as well as in Ref. \cite{carpentier-bglass-layered}).
This is also the valued we inferred from the static
and dynamic RG equations of
\cite{goldschmidt-houghton,goldschmidt-dynamics-co}
and we thus agree with their RG equations. This value
was $R=2 \pi$ and thus incorrect in \cite{hwa-fisher-co}
but later corrected back to $R=\pi$ \cite{hwa-privcomm}.

The origin of the 
discrepancy between (\ref{result-co}) and the result 
(5.24) of Ref. \cite{goldschmidt-houghton} thus lies
in the calculation of the correlation function. It can
probably be traced to the algebraic mistake between equations
(5.23) and (5.24) of Ref. \cite{goldschmidt-houghton}.
Finally the discrepancy with \cite{denis-bernard}
lies first in their RG equations with $N=1$ (we extracted
their amplitude ratio as being $R = \pi/2$) 
and later in a factor of two in the calculation of
the correlations (correcting for the additional misprint
between formulae (13) and (14)).

As discussed in the Introduction this improves the
comparison between numerical simulations and the RS-RG.
Indeed in Ref. \cite{numRSGM3} a result smaller
by a factor of $5$ than \cite{hwa-fisher-co} was found,
and we find here a result smaller than \cite{hwa-fisher-co}
by a factor of $4$. Concerning the recent 
quoted results \cite{rieger-zerot-co} of the numerical
determination of the ground state, we find that
it is about twice the naive continuation of our amplitude
to $T=0$,
i.e setting ($\tau=1$) in the above result (\ref{result-co}).

\section{Dynamics} \label{section4}

In this section we study the dynamics of the
triangular lattice on a disordered substrate.
We study directly the dynamics of the original
random sine gordon model (\ref{def-model}). The method we use
is to perform the renormalisation on the dynamical
effective action associated to (\ref{def-model}) and compute dynamical quantities.
We find that this method is more convenient, and leads to
more tractable calculations, than the 
method \cite{goldschmidt-dynamics-co,shapir-dynamics-co} originally
used by Goldschmidt and Schaub to study the dynamics
of the Cardy-Ostlund model.
In order to obtain the dynamical
exponent $z$ one needs,
in addition to purely dynamical quantities,
some information on the statics. It would be
potentially erroneous to attempt to extract
from the previous Section the necessary
information on the statics, 
since we used in the previous Section 
a different method with a different regularisation 
scheme. Instead, 
the consistent approach is
to use the same method and the same regularisation
scheme in the statics and the dynamics. We will
thus carefully reobtain the results for the
statics within the same dynamical RG. This is possible,
as we will demonstrate, because we are studying the 
{\it equilibrium} dynamics. Using the fluctuation dissipation  
theorem (FDT), which then holds exactly by assumption, one can then
reobtain the statics. We will also show (see Appendix \ref{appendixb}) 
that the same effective action method, but applied to the
replica symmetric theory gives the same results for the static
as the FDT equilibrium dynamics. Also
clearly demonstrate how the cutoff dependence comes 
in the RG equations and how at $T=T_g$ the
physical amplitudes become universal,
as detailed in Appendix \ref{appendixa}.

\subsection{perturbation theory on the dynamical action}
\label{section4a}

The dynamics of the model (\ref{def-model}) can be described by a 
Langevin type equation:

\begin{eqnarray}
\eta \frac{\partial}{\partial t} u^{\alpha}(x,t)
= - \frac{\delta H}{\delta u^{\alpha}(x,t)}
+ \zeta^{\alpha}(x,t)
\end{eqnarray}

with $\langle \zeta^{\alpha}(x,t) \zeta^{\beta}(x',t') \rangle = 2 \eta T 
\delta_{\alpha \beta} \delta(x-x') \delta(t-t')$
being the thermal noise and $\eta$ is the friction coefficient. The hamiltonian
$H$ is the sum $H_0+H_{dis}$ defined in (\ref{def-model}). The equation of motion reads, specifically:

\begin{eqnarray}
&& \eta \frac{\partial}{\partial t} u^{\alpha} = 
c_{66} \nabla^2 u^{\alpha} + (c_{11} - c_{66}) \partial_{\alpha} 
\partial_{\beta} u^{\beta}  \\ \nonumber
& & ~~~~~~~~~~~~+ f_1^{\alpha}(x) + f_2^{\alpha}(x)
+ \zeta^{\alpha}(x,t) \\
&& f_1^{\alpha}(x) = -  2 T \sqrt{g}  \sum_{\nu} \vec{K}^{\alpha}_{\nu}
\sin \left( \vec{K}_{\nu}. \vec{u}(x)+\phi_{\nu}(x) \right) \\
&& f_2^{\alpha}(x) = \frac{T}{2} ( \partial_{\beta} \sigma_{\alpha \beta}
+ \partial_{\beta} \sigma_{\beta \alpha } )
\end{eqnarray}

A convenient method to study the dynamics is to use
the de Dominicis-Janssen (or Martin Siggia Rose)
generating functional. Using the Ito prescription
it can be readily averaged over disorder. The disordered
averaged functional reads:

\begin{eqnarray}\label{action2}
&& Z[h,\hat{h}] = \int Du D\hat{u} e^{- S[u,\hat{u}] + h u + i \hat{h} \hat{u}} \\
&& S[u,\hat{u}] =   S_0[u,\hat{u}] + S_2[u,\hat{u}] + S_{int}[u,\hat{u}] \nonumber \\
&& S_0[u,\hat{u}] = \int_{q,t}
i \hat{u}^{\alpha}_{- q t}
( \eta \partial_t + c_{11} q^2 P_{\alpha \beta}^L
+ c_{66} q^2 P_{\alpha \beta}^T ) u^{\beta}_{q,t} \\ \nonumber
&& ~~~~~~~~~~~~~~~~~~~~~~~~
- \eta T \int_{r,t} (i \hat{u}^{\alpha}_{rt}) (i \hat{u}^{\alpha}_{rt}) \nonumber \\
&& S_2[u,\hat{u}]  =  - \frac{T}{2} \int_{q,t,t'}
(i \hat{u}^{\alpha}_{q,t}) (i \hat{u}^{\beta}_{-q,t'}) q^2 \\ \nonumber
&& ~~~~~~~~~~~~~~~~~~~~~~~~~~~~~~~~~~\times
(\Delta_{66} P_{\alpha \beta}^T(q) + 
\Delta_{11} P_{\alpha \beta}^L(q) )
\nonumber \\
&& S_{int}[u,\hat{u}]   =  - \frac{1}{2} \int_{r t t'} (i \hat{u}^{\alpha}_{rt})
(i \hat{u}^{\beta}_{rt'})
\Delta^{\alpha \beta}(u_{rt} - u_{rt'})
\end{eqnarray}

where the correlator of the pinning force is
$\overline{ f_1(r,u_{rt}) f_1(r',u_{r't'}) } =
\Delta^{\alpha \beta}(u_{rt} - u_{r't'})$. 
>From this functional one can obtain the
disorder averaged correlation function
$C^{\alpha,\beta}_{rt,r't'} = \overline{ \langle u^{\alpha}_{rt} u^{\beta}_{r't'} \rangle }$
and response function
$R^{\alpha,\beta}_{rt,r't'} = \delta \overline{ \langle u^{\alpha}_{rt} \rangle }/
\delta h^{\beta}_{r't'} $ which measures the response to a perturbation
applied at a previous time. They are obtained from the
above functional as 
$C^{\alpha \beta}_{rt,r't'} = \langle u^{\alpha}_{rt} u^{\beta}_{r't'} \rangle_S $
and $R^{\alpha \beta}_{rt,r't'} = \langle u^{\alpha}_{rt} i \hat{u}^{\beta}_{r't'} \rangle_S$
respectively. Causality imposes that $R_{rt,r't'}=0$ for $t'>t$
and the Ito prescription imposes that
$R_{rt,r't}=0$. All correlations $\hat{u} \hat{u}$ vanish.
We will assume here time and space translational invariance
and denote indifferently $C_{rt,r't'} =C_{r-r',t-t'}$ and
$R_{rt,r't'} =R_{r-r',t-t'}$ by the same symbol,
as well as their Fourier transforms when no confusion is possible.
Note that in this problem $C_{r,-t}=C_{r,t}$.

In the absence of disorder the action is simply quadratic
$S=S_0$ and the response function is thus
(for $t>0$): 

\begin{equation}
R^{\alpha \beta}_{q,t} = P^L_{\alpha \beta}(q) \mu e^{- c_{11} q^2 \mu t} \theta(t)
+ P^T_{\alpha \beta}(q) \mu e^{- c_{66} q^2 \mu t}  \theta(t)
\end{equation}

we have introduced the mobility $\mu = 1/\eta$. The 
correlation function is:

\begin{equation}
C^{\alpha \beta}_{q,t} = P^L_{\alpha \beta}(q) \frac{T}{c_{11} q^2} e^{- c_{11} q^2 \mu t} 
+ P^T_{\alpha \beta}(q) \frac{T}{c_{66} q^2} e^{- c_{66} q^2 \mu t}
\end{equation}

They satisfy the fluctuation dissipation theorem (FDT), i.e that:

\begin{equation} \label{fdt}
R^{\alpha \beta}_{r,t}
= - \theta(t) \frac{1}{T} \partial_t
C^{\alpha \beta}_{r,t}
\end{equation}

In the presence of disorder one will study perturbation theory
expanding in the interaction term $S_{int}$ using the quadratic
part $S_0 + S_2$ as the bare action. The disorder has a quadratic
part $S_2$ which is purely static and is immaterial in the
perturbation theory. Indeed the response function of
$S_0 + S_2$ is identical to the one of $S_0$ and 
the correlation function is changed as:
$C^{\alpha \beta}_{q,t} \to C^{\alpha \beta}_{q,t} +
{C_{stat}}^{\alpha \beta}_{q,t}$ with:

\begin{equation}
{C_{stat}}^{\alpha \beta}_{q,t} = T \frac{\Delta_{66}}{c_{66}^2 q^2}  P_{\alpha \beta}^T(q) + 
T \frac{\Delta_{11}}{c_{11}^2 q^2} P_{\alpha \beta}^L(q) 
\end{equation}

which is purely static and does not appear in any diagram of
perturbation theory. Thus for any practical purpose one can consider
that $S_0$ is used as the bare action.

We will perform the calculations on the
general form for the correlator of the disorder :

\begin{equation}
\Delta^{\alpha \beta}(u) = \sum_K \Delta^{\alpha \beta}_K e^{i K.u}
\end{equation}

In the case of the triangular lattice for later convenience the sum is over
the six reciprocal lattice vectors $K=1,2,3,-1,-2,-3$. 
We will specialize at the end to the
case of interest here, i.e a triangular lattice 
with the model introduced in Section II:

\begin{equation}
\Delta^{\alpha \beta}(u) = 2 g T^2 \sum_{K=1,2,3} K_{\alpha} K_{\beta} \cos(K.u)
\end{equation}
where we denote by $1,2,3$ the three reciprocal lattice vectors $K_{\nu}$
with $\nu = 1,2,3$ of (\ref{def-model}).

In fact, the condition $\Delta^{\alpha \beta}_K =  \Delta_K K_{\alpha} K_{\beta}$
is a potential condition which is preserved under RG and describes the
dynamics of the model studied in the previous Section.
With no loss of generality $\Delta(u)$ is an even function and thus
$\Delta_K = \Delta_{-K}$ and $\Delta_K$ is real.
The model of Section II is thus obtained for
$\Delta_K = g T^2$, and more generally the relation with the statics
studied in Appendix C is $\Delta_K =2 T^2 g_K$.

To establish the dynamical RG equations we will compute the
effective action in perturbation of $S_{int}$ using
$S_0 + S_2$ as the bare theory. The calculation to second order is
performed in the appendix and we will only quote the results.
Here we compute the effective action $\Gamma[u,\hat{u}]$ to lowest order in the
interacting part $S_{int}$:

\begin{eqnarray}
\Gamma[u,\hat{u}] &  = & S_0[u,\hat{u}] + S_2[u,\hat{u}] \\ \nonumber 
& & +
\langle S_{int}[u+\delta u, \hat{u} + \delta  \hat{u}]
 \rangle_{\delta u, \delta \hat{u}} + O(S_{int}^2)
\end{eqnarray}

where the averages over $\delta u$, $\delta \hat{u}$ are 
performed with the action $S_0 + S_2$.
The calculation gives:

\begin{eqnarray} \nonumber
\Gamma[u,\hat{u}] & = & S_0 + S_2 - 
\int_{r t t'} R^{\gamma \beta}_{rt,rt'} (i \hat{u}^{\alpha}_{rt})
\langle \Delta^{\alpha \beta ; \gamma}(u_{rt} - u_{rt'}) \rangle  \\ 
& & - \frac{1}{2} \int_{r t t'} (i \hat{u}^{\alpha}_{rt}) (i \hat{u}^{\beta}_{rt'})
\langle \Delta^{\alpha \beta}(u_{rt} - u_{rt'}) \rangle 
\end{eqnarray}

Here the symbol $\langle F[u]\rangle $ means $\langle F[u+ \delta u]\rangle _{\delta u}$
and we have used the symmetry $\Delta^{\alpha \beta} = \Delta^{\beta \alpha}$.
We denote 
$\Delta^{\alpha \beta ; \gamma}(u) = \partial_\gamma \Delta^{\alpha \beta}(u)$
etc..
This can be rewritten as:

\begin{eqnarray} \nonumber
\Gamma[u,\hat{u}] & = & S_0 + S_2  +
\int_{r t t'} (i \hat{u}^{\alpha}_{rt}) \Sigma(u_{rt} - u_{rt'}, t-t')
u^{\beta}_{rt'} \\ 
& &- \frac{1}{2} \int_{r t t'}
(i \hat{u}^{\alpha}_{rt}) (i \hat{u}^{\beta}_{rt'})
D^{\alpha \beta}(u_{rt} - u_{rt'}, t-t')
\end{eqnarray}

with:
\passage
\begin{eqnarray}
&& \Sigma^{\alpha \beta}(u_{rt} - u_{rt'}, t-t') = 
R^{\gamma \delta}_{rt,rt'} ( 
\langle \Delta^{\alpha \beta ; \gamma \delta}(u_{rt} - u_{rt'}) \rangle 
- \delta_{tt'} \int_{t''}
R^{\gamma \delta}_{rt,rt''}
\langle \Delta^{\alpha \beta ; \gamma \delta}(u_{rt} - u_{rt''}) \rangle ) \\
&& =- \sum_K \Delta^{\alpha \beta}_Ke ^{i K (u_{rt} - u_{rt'})}
( K.R_{0,t-t'}.K  e^{- K.(C_{0,0} - C_{0,t-t'}).K }
- \delta(t-t') \int_\tau K.R_{0,\tau}.K  e^{- K.(C_{0,0} - C_{0,\tau}).K } )\\
&& D^{\alpha \beta}(u_{rt} - u_{rt'}, t-t')
= \langle \Delta^{\alpha \beta}(u_{rt} - u_{rt'}) \rangle 
= \sum_K \Delta^{\alpha \beta}_K
e^{- K.(C_{0,0} - C_{0,t-t'}).K }
e^{i K (u_{rt} - u_{rt'})}
\end{eqnarray}

\retour
One can check that the exact FDT equalities are satisfied (for $t>t'$):

\begin{eqnarray}
\partial_{t'} D^{\alpha \beta}(u_{rt} - u_{rt'}, t-t')
= - T \Sigma^{\alpha \beta}(u_{rt} - u_{rt'}, t-t')
\end{eqnarray}

where the time derivative acts only on the explicit time
dependence (i.e second argument) of the function
\cite{footnote2}

In the above effective action to first order we 
will keep only the relevant terms, namely:

(i) the disorder, by definition:

\begin{eqnarray}
\Delta_R^{\alpha \beta} = D^{\alpha \beta}(u_{rt} - u_{rt'}, \infty ) 
\end{eqnarray}

(ii) the thermal noise, by definition:

\begin{eqnarray}
\delta(\eta T)_{\alpha \beta} = \frac{1}{2} \int_\tau ( D^{\alpha \beta}(0, \tau) 
- D^{\alpha \beta}(0, \infty) )
\end{eqnarray}
which will be in practice a diagonal tensor
$\delta(\eta T)_{\alpha \beta} = \delta(\eta T) \delta_{\alpha \beta}$.

(iii) the friction coefficient, by definition:

\begin{eqnarray}
\delta \eta_{\alpha \beta} = \lim_{\omega \to 0} i \partial_{\omega}
\Sigma^{\alpha \beta}(0, \omega) =
- \int_{\tau >0} \tau \Sigma^{\alpha \beta}(0, \tau) 
\end{eqnarray}

Let us note that in the same way that the disorder term $D$
generated in the effective action
can be split into a time persistent (i.e non local in time) operator 
(disorder (i))
which become relevant at $T_g$ and an operator local in time
(thermal noise (ii)), the term $\Sigma$ gives a local operator
(friction (iii)) and an additional time persistent kinetic term
(iv) not written above. In the present approach these terms are 
directly related via the FDT.
and we do not need to consider (iv) separately, though it may be
important in other cases.

Using the above FDT relation, integrating by parts 
and using the symmetry $C_{r,-t} = C_{r,t}$ one
immediately finds that $T \delta \eta_{\alpha \beta} =
\delta(\eta T)_{\alpha \beta}$, i.e that temperature
is not renormalized:

\begin{eqnarray}
\delta T = 0
\end{eqnarray}

The non-renormalization of temperature holds to all
orders and is guaranteed by the FDT relations.

Thus, to first order one has simply to consider
the disorder:

\begin{eqnarray}
\Delta^R_K= \Delta_K e^{- K.(C_{0,0}- C_{0,t=\infty}).K } 
\end{eqnarray}

and the correction to the friction:

\begin{eqnarray*}
\delta \eta_{\alpha \beta} = \frac{1}{T} \sum_K \Delta_K^{\alpha \beta}
\int_{0}^{+\infty} dt
&& ( e^{- K.(C_{0,0} - C_{0,t}).K } \\ 
&&- e^{- K.(C_{0,0} - C_{0,t=\infty}).K} )
\end{eqnarray*}

Using that $\Delta_K^{\alpha \beta} =  \Delta_K K_\alpha K_\beta$,
$\Delta_K = g T^2$ for the model of interest and 
isotropy one gets that $\delta \eta_{\alpha \beta} = \delta \eta \delta_{\alpha \beta}$
with:

\begin{eqnarray} \label{eta-first}
\delta \eta = \frac{g T}{2} \sum_K K^2
\int_{0}^{+\infty} dt
&& ( e^{- K.(C_{0,0} - C_{0,t}).K } \\
&&- e^{- K.(C_{0,0} - C_{0,t=\infty}).K} )
\end{eqnarray}

which becomes infrared divergent below $T_g$.
We now turn to the full result up to second order and 
the renormalization.

\subsection{RG equations to second order 
and calculation of the dynamical exponent}
\label{section4b}

The calculation of the effective action $\Gamma$ to second order
in $S_{int}$ is performed in the 
Appendix. We now summarize all the results up to second order,
specializing to the case of interest here, i.e model (\ref{def-model}) with
$\Delta_K = g T^2$. We have also introduced the dimensionless
disorder strength $\tilde{g} = g a^2$. We have specified a 
short-scale regularization, which we keep as general as possible
(see Appendix for all details):

\begin{eqnarray}
C_{ij}(\vec{r},t,a) = && T
\int {d^2 \vec{q} \over (2 \pi)^2} 
{ \phi(a q) \over q^2}
e^{i \vec{q} .\vec{r}} \\ \nonumber
& \times &
\left( c_{11}^{-1} P_{ij}^L 
e^{- c_{11} q^2 \mu t } +c_{66}^{-1} P_{ij}^T e^{- c_{66} q^2 \mu t } \right)
\end{eqnarray}
with $\phi(0)=1$.

One has:

\begin{eqnarray} \label{definition-b}
B_{ij} (r,t,a) & = & 2 ( C_{ij}(0,0,a) - C_{ij}(r,t,a) )\\
 & = &  B_{ij}(r/a,t/a^2,1)
\end{eqnarray}

With these definitions one finds:
\passage
\begin{mathletters}
\label{resultfinal}
\begin{eqnarray} 
\label{g-renorm}
&& g_R = 
e^{- \frac{1}{2} K.B(0,t=\infty,a).K } a^{-2} ( \tilde{g} +
\tilde{g}^2 a^{-2}  \int_{r} ( e^{K.B(r,0,a).K'}
- 2 e^{- \frac{1}{2} K.B(r,0,a).K} )) \\
\label{d66-renorm}
&& \Delta_{66,R} = \Delta_{66} + 
\frac{1}{2} \tilde{g}^2 a^{-4} T \sum_K \int_r (\frac{3}{8} K^2 r^2 - \frac{1}{4} (K.r)^2 )
e^{- K.B(r,0,a).K} \\
\label{d11-renorm}
&& \Delta_{11,R} = \Delta_{11} +
\frac{1}{2} \tilde{g}^2 a^{-4} T  \sum_K \int_r (\frac{1}{8} K^2 r^2 + \frac{1}{4} (K.r)^2 )
e^{- K.B(r,0,a).K} \\
\label{eta-renorm}
&& \eta_R = \eta + \frac{1}{2} T \tilde{g} a^{-2} \sum_K K^2 \int_{0}^{+\infty} dt
e^{- \frac{1}{2} K.B(0,t,a).K }
\end{eqnarray}
\end{mathletters}

\retour
In the first line $K' \neq K$. The last line is our previous 
first order result (\ref{eta-first}).
Studying the variation with respect to an infinitesimal change of
cutoff $a \to a' = e^l a$ leads to the following RG equations,
derived in detail in the Appendix:

\begin{mathletters}
 \label{rgfinal}
\begin{eqnarray}
\label{rg-d66}
&& \frac{d \Delta_{66}(l)}{dl} = A_{66}(\phi) T \tilde{g}(l)^2 \\
\label{rg-d11}
&& \frac{d \Delta_{11}(l)}{dl} = A_{11}(\phi) T \tilde{g}(l)^2  \\
\label{rg-g}
&& \frac{d \tilde{g}(l)}{dl} = (2 - \kappa_1 K^2 T) \tilde{g}(l)
+ \tilde{g}(l)^2 A_{g}(\phi) \\
\label{rg-eta}
&& \frac{d \eta (l)}{dl} = A_{\eta}(\phi) T \tilde{g}(l) 
\end{eqnarray}
\end{mathletters}

with $T_c = 2/\kappa_1 K^2$. The amplitudes $A_i(\phi)$ in
general depend on the details of the regularisation and
are computed in the Appendix. When evaluated at $T=T_g$ however,
they have a simple form:
\begin{mathletters}
\label{constantes}
\begin{eqnarray} 
&& A_{66} = \frac{3 \pi}{4} K^2 e^{8 \pi C(\phi)} (2 I_0(\alpha) - I_1(\alpha)) \\
&& A_{11} = \frac{3 \pi}{4} K^2 e^{8 \pi C(\phi)} (2 I_0(\alpha) + I_1(\alpha))  \\
&& A_g = 2 \pi e^{4 \pi C(\phi)} (I_0(\alpha) - 2 I_0(\alpha/2)) \\
&& A_{\eta} = \frac{3}{2} e^{\gamma + 4 \pi C(\phi)} K^2
\frac{1}{c_{66}} ( \frac{c_{66}}{c_{11}} )^{\frac{c_{66}}{c_{11}+c_{66}}} \eta
= B  \eta
\end{eqnarray}
\end{mathletters}
One sees that all cutoff dependence drops out of the universal
ratios determining the critical exponents.

The dynamical exponent $z$ below $T_g$ is determined from:

\begin{eqnarray}
\frac{d \ln \eta(l)}{d l} = T B  \tilde{g}(l) 
\end{eqnarray}

At the fixed point $\tilde{g}(l) = \tilde{g}^* = - 2 \tau /A_g$
with $\tau = (T-T_c)/T_c$ one has $\eta(L) \sim L^{z-2}$
where the dynamical exponent is:

\begin{eqnarray*}
z - 2 = -2 \tau \frac{T_g B}{A_g} =
\frac{12 e^{\gamma} \tau}{2 I_0(\alpha/2) - I_0(\alpha)}
( \frac{c_{66}}{c_{11}} )^{\frac{c_{66}}{c_{11} + c_{66}}}
\frac{c_{11}}{c_{11} + c_{66} }
\end{eqnarray*}

using $K^2 T_g = 8 \pi/( c_{11}^{-1} + c_{66}^{-1} )$. This 
yields our result for the dynamical exponent to lowest order
in $\tau$:

\begin{eqnarray} \label{theta}
\frac{z - 2}{\tau} = \theta = 3 e^{\gamma} 
\frac{ (2+ \alpha) (\frac{2-\alpha}{2+ \alpha})^{\frac{2-\alpha}{4}} }{
2 I_0(\alpha/2) - I_0(\alpha)}
\end{eqnarray}

with $\alpha(T_g) = \alpha_g = 2 (c_{11}-c_{66})/(c_{11}+c_{66})
= 2 (1+\sigma)/(3-\sigma)$ where $\sigma$ is the Poisson ratio
defined as usual as $\sigma = 1 - 2 c_{66}/c_{11} = \lambda/(\lambda + 2 \mu)$. See
fig.3 for a plot of this exponent $\theta$. 

One checks that in the case of the $N=1$ Cardy Ostlund model
one recovers previous result. Performing the substitutions
explained before equation (\ref{RG-co}), with 
$\tilde{g}^* = \tau/(2 \pi)$, $B^{CO} = e^\gamma/c$, $T^{CO}_g = 4 \pi c$,
one finds: 

\begin{eqnarray}
z^{CO} - 2 = \tilde{g}^* T^{CO}_g B^{CO} = 
2 e^{\gamma} \tau
\end{eqnarray}

which is the result of Goldschmidt and Schaub\cite{goldschmidt-dynamics-co,shapir-dynamics-co}
 (they use $\sqrt{c} = e^{\gamma}/2$).

One can obtain further dynamical quantities near the
transition. Explicitly integrating the RG equation
(\ref{solution-RG}) one has:

\begin{eqnarray}
\int_0^l dl \tilde{g}(l) = 
\frac{1}{B_g} \ln(1 + \chi (e^{2 \tau l}-1) )
\end{eqnarray}

This yields the diffusion coefficient $D=c/\eta$
at scale $L=a \ln l$. We denote $D_0=D(a)$ the bare diffusion 
coefficient $D_0=D(a)$. Exactly at $T_g$ one finds:

\begin{eqnarray}
\frac{D(L)}{D_0} = exp( - B T \int_0^l dl \tilde{g}(l) ) =
\frac{1}{(1 + \frac{\ln(L/a)}{\ln(\xi_0/a)})^\theta}
\end{eqnarray}

with $\xi_0 = a \exp(1/(g_0 B_g))$ is the characteristic length
at $T_g$ and the exponent $\theta = \lim_{\tau \to 0^{+}} (z-2)/\tau$ is given
by the formula (\ref{theta}) above. 

Above and near $T_g$ one has:

\begin{eqnarray} \label{diffusion}
\frac{D}{D_0} = \exp( - \theta \ln(1 + \frac{g_0 B_g}{2 |\tau|} ) )
\end{eqnarray}

Finally below $T_g$ one has:

\begin{eqnarray}
\frac{D}{D_0} = (1 + \frac{g_0 B_g}{2 \tau} (e^{2 \tau l} -1) )^{\theta}
\sim e^{(2 - z) l}
\end{eqnarray}

One can also study the velocity $v$ of the system in
response to a small force $f$. We 
follow the analysis of \cite{shapir-dynamics-co}
based on stopped RG arguments as in \cite{nozieres-gallet}.
Above $T_g$ one has $v \sim \mu f$ with
$\mu = D$ (Einstein relation), and $D$ was obtained
in (\ref{diffusion}). In the glass phase $T \leq T_g$
one iterates the RG until a scale $L \sim \sqrt{c/f}$ and
then match the RG to an asymptotic ohmic regime.
One thus obtains the $v-f$ characteristics as
$v = \mu(L = \sqrt{c/f}) f$. For $\tau > 0$ one has:

\begin{eqnarray}
v \sim \frac{\mu_0 
f}{(1 + (\frac{f_c}{f})^{\tau} - (\frac{f_c}{f_0})^{\tau})^{\frac{z-2}{2 \tau}}}
\end{eqnarray}

with $f_0 = c/a^2$ and $f_c$ is the {\it effective critical force}.
Since we work at $T>0$ there cannot be a true critical force,
but one can always define an effective critical force
\cite{blatter-vortex-review}. In the case of vortices
in superconductors it is gives the critical current $j_c$.
It is also possible to relate $f_c$
to the Larkin length $R_c = a (2 \tau/g_0 B_g)^{1/(2\tau)}$ which is
similar to $R_a$ in this single cosine model. The relation
is $f_c/f_0 = (a/R_c)^2$. Thus for weak disorder one has:

\begin{eqnarray}
v \sim \mu_0 f \left(\frac{f}{f_c}\right)^{\frac{z-2}{2}}
\end{eqnarray}

Exactly at the transition 
$\tau=0$ one finds:

\begin{eqnarray}
v \sim \mu_0 \frac{f}{(1 + 
\frac{ \ln( \frac{f_0}{f} )}{\ln( \frac{f_0}{f_c} }) )^{\theta}}
\end{eqnarray}

with $f_c = c \xi_0^{-2}$. 

Note that we have used near-equilibrium arguments
which assume the irrelevance of KPZ type terms
near the transition and ignores possible violations
of Einstein relation. A more detailed analysis goes
beyond the scope of this paper \cite{shapir-dynamics-co}.

\section{Conclusion}
\label{section5}

In this paper we have studied the problem of a triangular elastic
lattice on a disordered substrate, excluding dislocations.
This problem is interesting in relation to vortex lattices
in superconductors, friction of surfaces, magnetic bubbles.
We have constructed the $N=2$ components model necessary to
describe correctly the triangular lattice and we have
studied both the statics and the dynamics using 
several methods of renormalization. These methods have
yielded consistent results. We have studied several regularizations
and showed explicitely that the amplitude ratios which determine the exponents
become independent of the regularization procedure at $T_g$.

We have obtained that there is a glass phase for $T<T_g$
where disorder is perturbatively relevant and yields to
behaviour qualitatively similar, but quantitatively more complex
than the $N=1$ component version of this model studied so far.
We found that the glass phase is described by a
a plane of fixed points, parametrized by temperature and
the Poisson ratio. We obtained that a $u^2 \sim A_1 \ln^2 r$ growth of
the static displacements and computed the amplitude
$A$ to lowest order in $T-T_g$. We found that it is a universal function
of the Poisson ratio. We showed that the asymptotic behaviour
of the correlation function in the glass phase is isotropic
$u_T \sim u_L$, but also found a universal subdominant
anisotropy $u^2_T - u^2_L \sim A_2 \ln r$. These behaviours are reminiscent of the
behaviour for the Bragg glass in $d=4-\epsilon$
or in $d=3$ obtained from the GVM, with the difference
that $u^2 \sim A_d \ln r$ in that case and $u^2_T - u^2_L \sim cst$.

We have also studied the equilibrium dynamics of the model. We have
obtained the dynamical exponent $z$, which is also a function
of temperature and the Poisson ratio. This value is larger
than the one for the $N=1$ component model indicating that the
system is more glassy. We also computed the  
effective critical current in the glass phase and related it to
the Larkin length.

Finally, we also reexamined the statics of the $N=1$ Cardy-Ostlund model
(planar random field XY model). We obtained an
expression for the amplitude of the $u^2 \sim A \ln^2 r$ 
growth of the displacements which seems compatible in order of
magnitude with the numerical simulations, though more extensive
simulations would be useful. Another prediction is that the
distribution of rescaled displacements $u/\ln r$ is a gaussian at large scale.
We propose that this could provide a new and non trivial numerical check of
the validity of the replica symmetric Cardy-Ostlund RG.

We thank T. Giamarchi, L. Cugliandolo, J. Kurchan
for useful discussions.

{\it Note added}:

While this manuscript was under completion we received a
preprint by C. Carraro and D.R. Nelson, cond-mat/9607184
who study the same model (\ref{def-model}). Their results
are different from ours. Their static RG calculation 
does not take into account angular integrals and
thus shows no dependence
on the Poisson ratio. The two calculations may coincide
only for the case $\alpha=0$, but a direct comparison
was difficult since they did not compute 
the $\ln^2 r$ amplitude. Our result for $z$, obtained
by carefully using the same 
regularization in the statics and the dynamics,
also disagrees with their result.

\appendix

\section{regularisation and universality}
\label{appendixa}
 In this appendix we detail
the method we used to renormalise eqs. (\ref{resultfinal}), with a general regularisation
 function $\phi$. This is interesting for several reasons: first as
we have seen it is crucial to use 
the same cut-off procedure in the dynamics and static RG approach. Second, using a general cut-off 
gives a well-defined method to determine the universal quantities, and moreover 
it might be crucial 
to avoid the usual logarithmic approximation when comparing results with experiments (or numerical calculations) :
 for the case of $He$ film, see the discussion of Nozi\`eres and Gallet\cite{nozieres-gallet}. 
   The procedure we develop originates from the momentum renormalisation of the XY model by 
Knops and Den Ouden\cite{knops-regularisation}. We use here a method that allows a 
real-space renormalisation : thus direct comparison with the hard core procedure is possible.

In order to study a change of the cutoff in eqs. (\ref{resultfinal}) we introduce:

\begin{eqnarray}
 h_{ij}(\vec{r},t,a) =&&  a T \int { d^2 \vec{q} \over (2\pi)^2}
 {\phi'(aq) \over q} e^{i \vec{q}.\vec{r}} \\ \nonumber
&& \times
\left( c_{11}^{-1} P_{ij}^L 
e^{- c_{11} q^2 \mu t } +c_{66}^{-1} P_{ij}^T e^{- c_{66} q^2 \mu t } \right)
\end{eqnarray}

Note that the above integral for $h_{ij}(\vec{r},t,a)$ 
has no infrared divergence in $q$. First we notice that (with $\kappa_1$ defined in (\ref{CG-interaction})) :
$$K_i h_{ij}(\vec{0},0,a) K'_j = -\kappa_1 T K.K' $$

and also:

\begin{eqnarray}
\label{prop-h}
 h_{ij}(\vec{r},t,a)  & = & h_{ij}(0,0,a) - \frac{1}{2} a d/da B_{ij}(\vec{r},t,a)\\
  & = &  h_{ij}(0,0,a) + \frac{1}{2} \vec{r}. \nabla_{\vec{r}} B_{ij}(\vec{r},0,a)\\
& = &  h_{ij}(0,0,a) + t \partial_t B_{ij}(\vec{0},t,a)
\end{eqnarray}

Then upon a change of the cut-off parameter 
$a\rightarrow \tilde{a}=ae^l$, we have the following change of cut-off function 
to first order in $l$:
 
\begin{equation}
\phi(aq)= \phi(\tilde{a}q) -  q l \tilde{a} \phi'(\tilde{a}q)
\end{equation}

This gives:

\begin{eqnarray}
\label{B-rescaled}
B_{ij} (r,t,a) = B_{ij} (r,t,\tilde{a}) + 2 l (h_{ij}(r,t,\tilde{a})
- h_{ij}(0,0,\tilde{a}) )
\end{eqnarray}

 Let us focus for example on the definition of the renormalised 
$\Delta_{66,R}$ (eq. \ref{d66-renorm}). Using (\ref{B-rescaled}) 
in it we obtain :

\begin{eqnarray}
\Delta_{66,R} & = & \Delta_{66} + 
\frac{1}{2} \frac{\tilde{g}^2}{\tilde{a}^4} T
e^{2 l (2 - T \kappa_1 K^2)} 
\sum_{\vec{K}} \int_r  e^{- K.B(r,0,\tilde{a}).K} \\ \nonumber
&&\times 
\left(\frac{3}{8} K^2 r^2 - \frac{1}{4} (K.r)^2 \right)
(1 - 2 l ~K.h(r,0,\tilde{a}).K) 
\end{eqnarray}

 which can be put in the same form as the original
equation (\ref{d66-renorm})
with the following definition of the running coupling constants:

\begin{eqnarray}
&& \tilde{g}(l) = \tilde{g} e^{ l (2 - T \kappa_1 K^2)} \\
&& \frac{d \Delta_{66}(l)}{dl} = A_{66}(\phi) T \tilde{g}(l)^2 
\end{eqnarray}

The scaling amplitude $A_{66}(\phi)$ depends on the cutoff function
but not on $\tilde{a}$ which can be eliminated by rescaling
$r/\tilde{a} \to r$  :

\begin{eqnarray}
\label{def-a-66}
A_{66}(\phi) = &&
- \sum_{\vec{K}} \int_{\vec{r}} \left(\frac{3}{8} K^2 r^2 - \frac{1}{4} (K.r)^2 \right)\\ \nonumber
&& ~~~~~\times
K.h(r,0,1).K  e^{- K.B(r,0,1).K} 
\end{eqnarray}

Note that this integral over $r$ is by construction finite and
convergent in the infrared and ultraviolet.

Similarly from (\ref{d11-renorm},\ref{eta-renorm}) 
one obtains the equations (\ref{rg-d11},\ref{rg-eta}) with the RG amplitudes 

\begin{eqnarray}
\label{def-a-11}
&& A_{11}(\phi) =
- \sum_{\vec{K}} \int_{\vec{r}} \left(\frac{1}{8} K^2 r^2 + \frac{1}{4} (K.r)^2 \right)\\ \nonumber
&& ~~~~~\times K.h(r,0,1).K  e^{- K.B(r,0,1).K}\\
\label{def-a-eta}
&& A_{\eta}(\phi) =
- \sum_{\vec{K}} K^2 \int_0^{+ \infty} dt\\ \nonumber
&& ~~~~~\times K.h(0,t,1).K  e^{- \frac{1}{2} K.B(0,t,1).K}
\end{eqnarray}

The same method can be used for the RG equation of $\tilde{g}$ in the case $t=\infty$:

\begin{eqnarray}
B_{ij}(0,\infty, a) & = & B_{ij}(0, \infty, \tilde{a}) 
+ 2 l (h_{ij}(r,\infty,\tilde{a})
- h_{ij}(0,0,\tilde{a}) )\\
& = & B_{ij}(0, \infty, \tilde{a})  +2l T \kappa_1 \delta_{ij}
\end{eqnarray}

We used  $h_{ij}(r,t= \infty,\tilde{a})=0$. Thus we obtain the RG eq. (\ref{rg-g}) with 

\begin{eqnarray}
\label{def-a-g}
 A_{g}(\phi) = 2 \int_{\vec{r}} && 
( K.h(r,0,1).K' e^{K.B(r,0,1).K'} \\ \nonumber
&&+ K.h(r,0,1).K e^{- \frac{1}{2} K.B(r,0,1).K}  )
\end{eqnarray}

 As we have seen these RG amplitudes depends in a non trivial way on the cut-off procedure 
(here the fonction $\phi$). However they enter (for example) the expression of the 2-point 
correlation functions, whose asymptotic behaviour should be universal at $T_c$. Thus we have to 
compute the values of the amplitudes $A_i(\phi)$ at $T=T_c$. As we will see they involves the asymptotics 
of the propagators $B_{ij}(\vec{r},0,a) \ (|\vec{r}| \gg a)$ and 
$B_{ij}(0,t,a) \ (t \gg a^2)$. 

First we focus on the $r$-propagator, which gives also 
the asymptotic interaction of the Coulomg-Gas (see (\ref{CG-interaction})):
 $$V(\vec{r} ,a)=B_{ij}(\vec{r},0,a)= E(r) \delta_{ij} + F(r) \hat{r}_i \hat{r}_j$$. 

 where $F(r)$ is given by 

\begin{eqnarray}
F(r)  & = &  2 T {1 \over 2 \pi} ( c_{11}^{-1}-c_{66}^{-1} )\int dq { \phi(aq) \over q}
 J_2(qr)
\end{eqnarray}

The other coefficient is given by  $ 2 E(r) = Tr.B(r) - F(r)$ where 

\begin{eqnarray}
 Tr.B(\vec{r})& = & \sum_i B_{ii}(\vec{r},0,a) \\ \nonumber& = & 
  {T \over  \pi} ( c_{11}^{-1}+c_{66}^{-1} )
\int { d^2 \vec{r}'\over a^2} \phi(r'/ a)
\ln \left( {|\vec{r}-\vec{r}'| \over |\vec{r}'| } \right)
\end{eqnarray}

 The large $r$ behaviour of the integral follows from the formula ($ |\vec{r}| \gg a$):

\begin{eqnarray} \label{formul-int-ang}
 \int {d^2 \vec{q} \over (2 \pi)^2} { \phi(a q) \over q^2}
\left( e^{i \vec{q}.\vec{r}} - 1 \right)
& \approx & -{1 \over 2 \pi}  \ln (r/a) + C(\phi)  \\
\int dq { \phi(aq) \over q}
 J_2(qr)  \approx \frac{1}{2} \phi(0)  
\end{eqnarray}
where $C(\phi) = {1 \over 2 \pi}\int d^2 \vec{r}'\phi(r')\ln (r') $.
 The asymptotics of the interaction is then :

\begin{eqnarray}
\label{real-asympt1}
E(r) &\approx  & 2 T \kappa_1 \left[ \ln(r/a) - 2 \pi C(\phi) \right] + T \kappa_2 \\
\label{real-asympt2}
 D(r) &\approx  &  - 2 T \kappa_2
\end{eqnarray}

 Now we turn back to the time propagator, which is given by 

\begin{equation}
\label{time-B-asymp}
 B_{ij}(0,t,a) = T \delta_{ij} \left[ c_{11}^{-1} F(c_{11}, t,a) + c_{66}^{-1} F(c_{66}, t,a) \right]
\end{equation}
where we define the function 

\begin{eqnarray}
F(c,t,a) & = &  \int {d^2 \vec{q} \over (2 \pi)^2} { \phi(aq) \over q^2}
\left( 1-e^{-c q^2 t \mu } \right)\\
& = & - {1 \over 4 \pi} \int d^2 \vec{r} \phi(r)
Ei \left( - {r^2 a^2\over 4 c t } \right)
\end{eqnarray}
where $Ei$ means the Exponential-Integral function which verifies ($\gamma$ is the Euler constant) :

$$Ei(-x) = \gamma + \ln(x) + \sum_{n=1}^{\infty} {a_n \over x^n} $$ 

 Taking the limit of large time in (\ref{time-B-asymp}) using the above equation, we find 

\begin{eqnarray}
\label{time-prop-asympt}
B_{ij}(0,t,a)\approx && T \delta_{ij} \left[ \kappa_1 \left( \ln \left( {4 t \mu \over a^2}\right)
- \gamma  - 4 \pi C(\phi) \right) \right.\\ \nonumber
&& \left. + {1 \over 4 \pi} \left( c_{11}^{-1} \log c_{11} 
+ c_{66}^{-1} \log c_{66} \right) \right]
\end{eqnarray}

 We are now able to find the expression of the RG amplitudes $A_i(\phi)$ at $T_c$. We will give a detailed 
calculation for $A_{\eta}(\phi)$, the others being similar. Using 
the property (\ref{prop-h}) : $K.h(0,t,1).K = -2 T/T_c + t \partial_t (K.B(0,t,1).K)$, we can write 
$A_{\eta}(\phi)$ (eq. (\ref{def-a-eta})) (after an integration by part) as 

\begin{eqnarray}
A_{\eta}(\phi) & = & 
 - \frac{1}{2} \sum_{\vec{K}} \left(  
2(1-{T \over T_c}) \int_0^{\infty} dt e^{-{1 \over 2} K.B(0,t,1).K} \right.\\ \nonumber
&& ~~~~~~~~~~~~\left.
- 2 \left[ t e^{-{1 \over 2} K.B(0,t,1).K}\right]_0^{\infty} \right)
\end{eqnarray}

This expression is valid for $T>T_c$,as well as for $T=T_c$
provided the first term is set to 0.
With the asymptotic form of the time correlator (\ref{time-prop-asympt}) :
$$ 
e^{-{1 \over 2} K.B(0,t,1).K} \simeq 
(4t \mu )^{-{T \over T_c}} \Gamma_K $$

with 
\begin{eqnarray}
 \Gamma_K = &&
\exp\left(  -{1\over 2} T K^2 \left( - \kappa_1
(\gamma  + 4 \pi C(\phi)) 
 \right. \right.\\ \nonumber
&&  \left. \left.
+ {1 \over 4 \pi} ( c_{11}^{-1} \log c_{11} 
+ c_{66}^{-1} \log c_{66} ) \right)\right)
\end{eqnarray}

we obtain the expression at $T_c$ :
\begin{equation}
A_{\eta} = \frac{1}{4} \sum_{\vec{K}} K^2 \mu^{-1} \Gamma_K \ \ \ (T=T_c)
\end{equation}

 We now turn back to the other amplitudes (\ref{def-a-11},\ref{def-a-66},\ref{def-a-g}). 
By the same method, and using the asymptotics (\ref{real-asympt2}) :

\begin{equation}
\label{asympt-r-1}
 e^{-K.B.K} \simeq r^{-4 {T \over T_c}} e^{2 \alpha (\hat{K}.\hat{r})^2 } e^{-\alpha + 8 \pi C(\phi)} 
\end{equation}

 where $\alpha$ has been defined in section II : $\alpha = \kappa_2 K^2 T$ we obtain at $T=T_c$:
\begin{eqnarray}
A_{66}(\phi) & = &   {1 \over 2} e^{-\alpha + 8 \pi C(\phi)} \\ \nonumber 
&& \times
\left( 
\sum_{\vec{K}} \int_0^{2 \pi} K^2  d\theta \left(\frac{3}{8}   - \frac{1}{4} \cos^2 (\theta) \right)
e^{2 \alpha \cos^2 (\theta) } \right) \\ \nonumber 
A_{11}(\phi) & = &  {1 \over 2} e^{-\alpha + 8 \pi C(\phi)}  \\ \nonumber 
&& \times
\left( 
\sum_{\vec{K}} \int_0^{2 \pi}  K^2 d\theta \left(\frac{1}{8}  + \frac{1}{4} \cos^2 (\theta) \right)
e^{2 \alpha \cos^2 (\theta) } \right)
\end{eqnarray}

For $A_g$ the integral split into two terms and we need another asymptotic expression (the first one follows 
directly from (\ref{asympt-r-1})):

$$ e^{{1 \over 2}K.B.K'} \simeq r^{-2 {T \over T_c}} e^{-{\alpha \over 2} + 4 \pi C(\phi)} 
e^{- 2\alpha (\hat{K}.\hat{r})(\hat{K}'.\hat{r}) } $$

Thus 
\begin{eqnarray}
A_g & = &  
2 e^{-{\alpha \over 2} + 4 \pi C(\phi)} \\ \nonumber
& & \times
\int_0^{2 \pi} d \theta 
\left( {1 \over 2} e^{- 2\alpha \cos(\theta) \cos(\theta+2\pi/3) } -e^{- \alpha \cos^2(\theta)} \right)
\end{eqnarray}

Performing the angular integration one finds the formulae (\ref{constantes})
in the text.

\section{static RG calculation from the effective action}
\label{appendixb}

In this Appendix we obtain the perturbation theory
result (\ref{resultfinal}) from the statics replica
calculation, using the effective action method.

The replicated partition function is $Z_n = Tr_n e^{-S}$
with $S[u] = S_0[u] + S_{int}[u]$ and:

\begin{eqnarray}\label{action3}
S_0[u] &=& \sum_{ab} \frac{1}{2} (C^{-1})^{ab,\alpha \beta}(q)
u_{a,\alpha}(-q) u_{b,\beta}(q)  \nonumber \\
S_{int}[u] &=& - \int_r \sum_{ab} R(u^{a}_r-u^{b}_r)
\end{eqnarray}

We will study the general model:

\begin{equation}
\label{model-general}
R(u)= \sum_K g_K e^{i K.u}
\end{equation}

In the case
of the triangular lattice, 
the sum is over the six reciprocal lattice vectors
$K=1,2,3,-1,-2,-3$. With no loss of generality $R(u)$ is an even function and thus
$g_K=g_{-K}$ and $g_K$ is real. Note here the redundancy in the sum under the
symmetry ($K \leftrightarrow -K$, 
$(a,b) \leftrightarrow (b,a)$).

The original model (\ref{def-model}) we want to study
is:

\begin{equation}
R(u) = g \sum_{K=1,2,3} \cos(K.u)
\end{equation}

where we denote by $1,2,3$ three reciprocal lattice vectors.
This model is obtained as a particular case of (\ref{model-general}) for $g_K = g/2$.

To obtain the effective action $\Gamma[u]$ to
second order in disorder, one has to study all possible Wick contractions of
the two vertex operators, with all one particle reducible diagrams being
absent (see below):

\begin{equation}
- \frac{1}{2} \int_{rr'} \sum_{abcd} R(u^{a}_r-u^{b}_{r} + \delta u^{a}_r - \delta u^{b}_r )
R(u^{c}_{r'}-u^{d}_{r'} + \delta u^{c}_{r'} - \delta u^{d}_{r'} )
\end{equation}
 
The contractions are of the form:

\begin{eqnarray}
-2 \Gamma^{(2)} & = & \langle R(z_1) R(z_2) \rangle  - \langle R(z_1) \rangle  \langle R(z_2) \rangle  \\ \nonumber 
 & & - \langle R'(z_1) \rangle  \langle z_1 z_2\rangle  \langle R'(z_2) \rangle 
\end{eqnarray}

In Fourier it reads:

\begin{eqnarray}
\int_{q_1, q_2} && 
R(q_1) R(q_2) e^{-\frac{1}{2} q_1^2 \langle z_1^2\rangle } e^{-\frac{1}{2} q_2^2 \langle z_2^2\rangle }
\\ \nonumber 
& & \times
(e^{-q_1 q_2 \langle z_1 z_2\rangle} - 1 + q_1 q_2 \langle z_1 z_2\rangle)
\end{eqnarray}

The result is:

\begin{equation}
\Gamma_2 = - \frac{1}{2} \int_{rr'} \sum_{abcd} \sum_{K K'} V^{ab,cd}_{K, K'}(r-r')
e^{i K. (u^{a}_r-u^{b}_{r})} e^{i K'. (u^{c}_{r'}-u^{d}_{r'})}
\end{equation}

with

\begin{eqnarray}
 V^{ab,cd}_{K K'}(r) & = & g_K g_{K'} e^{-\frac{1}{2} K.B^{ab}(0).K}
e^{-\frac{1}{2} K'.B^{cd}(0).K'} \\ \nonumber 
& & \times 
( e^{ - K.(C^{ca}_r + C^{db}_r - C^{cb}_r - C^{da}_r).K' } - 1  
\nonumber \\
&& + K.(C^{ca}_r + C^{db}_r - C^{cb}_r - C^{da}_r).K' )
\end{eqnarray}

The points $r$ and $r'$ are constrained to be within a small
distance. The contraction of the points $r$ and $r'$ result
in new local operators which will correct the effective action. 
The coefficients can be computed using a standard
short distance expansion. The relevant new terms are
of two types:

(i) {\it the fusion terms}: 

they are of the form:

\begin{eqnarray}
\Gamma_2 = \int_{r} \sum_{ab} \sum_{K} \delta g_K e^{i K.(u^a_r - u^b_r)}
\end{eqnarray}

These fusion terms can be split in two types,
$\delta g_K= \delta_1 g_K + \delta_2 g_K$, the
same $K$ vector $\delta_1 g_K$ and different 
$K$ vector $\delta_2 g_K$.

(i1) {\it same $K$ fusion terms}: 

The terms ( $K'=K$, $c=b$ or $d=a$) and
($K'=-K$ and $c=a$ or $d=b$) lead to:

\begin{eqnarray}
\delta_1 g_K = \frac{1}{2} \int_r \sum_c &&
 ( V^{ac,cb}_{K, K}(r) + V^{ac,bc}_{K, - K}(r) \\ \nonumber
& & +
V^{cb,ac}_{K, K}(r) + V^{cb,ca}_{K, - K}(r) )
\end{eqnarray}

(there are of course four other terms obtained by the
symmetry ($K \leftrightarrow -K$, 
$(a,b) \leftrightarrow (b,a)$) which correct the
corresponding symmetric term).

This correction can then be computed explicitly.
One finds that for $a \neq b$ performing a sum over $c \neq a,b$:

\begin{equation}
\delta_1 g_K = 2 (n-2) e^{ - \frac{1}{2} K.B^{ab}(0).K } (g_K)^2
\int_r e^{ - \frac{1}{2} K.B(r).K } A(r)
\end{equation}

with

\begin{equation}
A(r) = 1 - e^{- K.C(r).K } (1 + K.C(r).K ) 
\end{equation}

This yields, upon setting $g_K = g/2$, part of the equation (\ref{resultfinal}).

(i2) {\it different $K$ fusion terms}:

These terms are a peculiarity of the triangular lattice.
This other correction comes from $a=b$ and $c=d$, or $a=d$ and $c=b$.

\begin{equation}
\delta_2 g_{K+K'} = \frac{1}{2} \int_r ( V^{ab,ab}_{K, K'}(r)  + V^{ab,ba}_{K, - K'}(r) )
\end{equation}

For a given $K_3$, for each $(K_1,K_2)$ such that $K_1 + K_2 = K_3$ one finds:

\begin{equation}
\delta_2 g_{K_3} = g_{K_1} g_{K_2} e^{- \frac{1}{2} K_3.B^{ab}(0).K_3 }
\int_r e^{ K_1.B(r).K_2 }
\end{equation}

This yields, upon setting $g_K = g/2$, the remaining 
part of the equation (\ref{resultfinal}),
taking into account 
that two couples $(K_1,K_2)$ and $(K_2,K_1)$ correct $g_{K_3}$.

(ii) {\it annihilation terms}:

Finally the subcase $K'=-K$ gives a correction to the quadratic
hamiltonian. It gives the local operator:

\begin{eqnarray}
\Gamma_2 & = & (- \frac{1}{2}) \int_r \sum_{ab} \sum_K
(- \frac{1}{2}) \\ \nonumber
 & & ~~~\times [ (r.\nabla) (u^a - u^b).K ]^2 
( V^{ab,ab}_{K, - K}(r)  + V^{ab,ba}_{K, K}(r) )
\end{eqnarray}

which gives:

\begin{equation}
\Gamma_2 = -
\partial_\alpha u^a_\gamma \partial_\beta u^b_\delta
\sum_K (g_K)^2 K_\gamma K_\delta \int_r r_\alpha r_\beta 
e^{-K. B(r).K}
\end{equation}

This should correct:

\begin{equation}
- \frac{1}{2 T} \partial_\alpha u^a_\gamma \partial_\beta u^b_\delta
( (\Delta_{11} - \Delta_{66}) \delta_{\alpha \gamma} \delta_{\beta \delta}
+ \Delta_{66} (\delta_{\alpha \beta} \delta_{\gamma \delta} ) )
\end{equation}

Let us define a tensor:

\begin{equation}
T_{\alpha \gamma, \beta \delta} = A \delta_{\alpha \beta} \delta_{\gamma \delta}
+ \frac{B}{2} (\delta_{\alpha \gamma} \delta_{\beta \delta} + 
\delta_{\alpha \delta} \delta_{\beta \gamma} )
\end{equation}

Then one finds:

\begin{equation}
A = \frac{3}{8} T_{\alpha \gamma, \alpha \gamma} - \frac{1}{4} 
T_{\alpha \alpha, \gamma \gamma}~~ 
B=  - \frac{1}{4} T_{\alpha \gamma, \alpha \gamma}
+ \frac{1}{2}  T_{\alpha \alpha, \gamma \gamma}
\end{equation}

Using this algebra it yields, upon setting $g_K = g/2$, the
corresponding part of the equation (\ref{resultfinal}).

\unecol

\section{dynamical effective action to second order}
\label{appendixc}

In this appendix we derive the perturbative expression
of the effective dynamical action 
to second order in disorder. We then identify the relevant
terms which correct the bare disorder and lead to 
divergences below $T_g$. This procedure amounts to 
a short distance, short time operator product expansion.
Note that the operators are local in $r$ but non local in time,
which makes the expansion more involved.

The effective action to
second order in the interaction term is \cite{goldschmidt-dynamics-co}:

\begin{equation}
-2 \Gamma^{(2)}[U] =  \langle S_{int}[U+\delta U]^2  \rangle_{\delta U}
-   \langle S_{int}[U+\delta U] \rangle^2_{\delta U} -
\langle \frac{\delta S_{int}[U+\delta U]}{\delta U} \rangle_{\delta U}  G
\langle \frac{\delta S_{int}[U+\delta U]}{\delta U} \rangle_{\delta U}
\end{equation}

with $U=(u,\hat{u})$ and $\delta U=(\delta u,\delta \hat{u})$
and a gaussian average over $\delta U$ is performed using 
the bare quadratic action $S_0 + S_2$ in (\ref{action2}).
The last term merely ensures
that all one particle reducible diagrams be absent.

One thus has to study all possible Wick contractions of
the two vertex operators:

\begin{equation}
i \hat{u}^{\alpha_1}_{r_1 t_1} i \hat{u}^{\beta_1}_{r_1,t'_1} 
\Delta_{\alpha_1 \beta_1}(u_{r_1,t_1} - u_{r_1,t'_1})
\Delta_{\alpha_2 \beta_2}(u_{r_2 t_2} - u_{r_2,t'_2})
i \hat{u}^{\alpha_2}_{r_2 t_2} i \hat{u}^{\beta_2}_{r_2,t'_2}
\end{equation}

imposing that at least two contractions joining the two vertices 1 and 2.

A tedious calculation gives, for the $i \hat{u} i \hat{u}$ term
(in short notations):

\begin{eqnarray}
&& -2 \Gamma =
2 (i \hat{u}^{\alpha_1}_{t_1}) (i \hat{u}^{\alpha_2}_{t_2})
\langle \Delta_{\alpha_1 \beta_1} (u_{t_1} - u_{t'_1}) 
\Delta_{\alpha_2 \beta_2; \gamma \delta} (u_{t_2} - u_{t'_2}) \rangle_c
R^{\delta \beta_2}_{t_2 t'_2}
( R^{\gamma \beta_1}_{t_2 t'_1} - R^{\gamma \beta_1}_{t'_2 t'_1} )
\nonumber \\
&& +
\frac{1}{2} (i \hat{u}^{\alpha_2}_{t_2}) (i \hat{u}^{\beta_2}_{t'_2})
\langle \Delta_{\alpha_1 \beta_1} (u_{t_1} - u_{t'_1}) 
\Delta_{\alpha_2 \beta_2; \gamma \delta} (u_{t_2} - u_{t'_2}) \rangle
(  R^{\gamma \alpha_1}_{t_2 t_1} - R^{\gamma \alpha_1}_{t'_2 t_1} )
(  R^{\delta \beta_1}_{t_2 t'_1} - R^{\delta \beta_1}_{t'_2 t'_1} )
+
\nonumber \\
&&
(i \hat{u}^{\alpha_1}_{t_1}) (i \hat{u}^{\alpha_2}_{t_2})
\langle \Delta_{\alpha_1 \beta_1; \delta} (u_{t_1} - u_{t'_1}) 
\Delta_{\alpha_2 \beta_2; \gamma} (u_{t_2} - u_{t'_2}) \rangle
( R^{\gamma \beta_1}_{t_2 t'_1} - R^{\gamma \beta_1}_{t'_2 t'_1} )
( R^{\delta \beta_2}_{t_1 t'_2} - R^{\delta \beta_2}_{t'_1 t'_2} )
\nonumber \\
&& +
\langle\Delta_{\alpha_1 \beta_1; \delta} (u_{t_1} - u_{t'_1}) 
\Delta_{\alpha_2 \beta_2; \gamma} (u_{t_2} - u_{t'_2})\rangle_c
(i \hat{u}^{\alpha_1}_{t_1}) (i \hat{u}^{\beta_1}_{t'_1})
R^{\gamma \beta_2}_{t_2 t'_2}
( R^{\delta \alpha_2}_{t_1 t_2} - R^{\delta \alpha_2}_{t'_1 t_2} )
\end{eqnarray}

Using the assumption of time and space translational invariance it can be put under
the form $\Gamma =
- \frac{1}{2} (i \hat{u}^{\alpha}_{r t_1}) (i \hat{u}^{\beta}_{r+r',t_2})
\delta \Delta^{\alpha \beta}_{r'}$. One finds
four terms:
$\delta \Delta = \sum_{i=1,4} \delta \Delta^{(i)}_{eff}$:

\begin{eqnarray} \label{terms}
\delta \Delta^{(1)}_{r'} &=& 
2 R^{\delta \lambda}(\tau_2,0) 
R^{\gamma \rho}(\tau_1,r')
\langle \Delta_{\beta \lambda; \gamma \delta} (u_{r+r', t_2} - u_{r+r', t_2-\tau_2})
\nonumber \\
&&
[ \Delta_{\alpha \rho} (u_{r t_1} - u_{r, t_2 -\tau_1}) 
- \Delta_{\alpha \rho} (u_{r t_1} - u_{r, t_2 -\tau_1 - \tau_2}) ] \rangle_c
\nonumber \\
\delta \Delta^{(2)}_{r''} &=&
\frac{1}{2} \delta(r'') R^{\gamma \rho}(\tau,- r')
R^{\delta \lambda}(\tau',- r')
\langle \Delta_{\alpha \beta ; \gamma \delta} (u_{r t_1} - u_{r t_2})
\nonumber \\
&&
[ \Delta_{\rho \lambda}(u_{r+r', t_1-\tau} - u_{r+r', t_1-\tau'}) +
\Delta_{\rho \lambda}(u_{r+r', t_2-\tau} - u_{r+r', t_2-\tau'}) -
\nonumber \\
&&
\Delta_{\rho \lambda}(u_{r+r', t_1-\tau} - u_{r+r', t_2-\tau'}) -
\Delta_{\rho \lambda}(u_{r+r', t_2-\tau} - u_{r+r', t_1-\tau'}) ] \rangle
\nonumber \\
\delta \Delta^{(3)}_{r'} &=& 
R^{\gamma \rho}(\tau_2, r')
R^{\delta \lambda}(\tau_1,- r') [ 
\langle \Delta_{\alpha \rho; \delta}(u_{r,t_1} - u_{r, t_2 - \tau_2})
\Delta_{\beta \lambda; \gamma}(u_{r+r',t_2} - u_{r+r', t_1 - \tau_1}) \rangle -
\nonumber \\
&&
\langle \Delta_{\alpha \rho; \delta}(u_{r,t_1} - u_{r, t_1 -\tau_1 - \tau_2})
\Delta_{\beta \lambda; \gamma}(u_{r+r',t_2} - u_{r+r', t_1 - \tau_1}) \rangle -
 \nonumber \\
&& \Delta_{\alpha \rho; \delta}(u_{r,t_1} - u_{r, t_2 - \tau_2}) 
\Delta_{\beta \lambda; \gamma}(u_{r+r',t_2} - u_{r+r', t_2 - \tau_1 - \tau_2}) \rangle ]
\nonumber \\
\delta \Delta^{(4)}_{r''} &=& \delta(r'')
R^{\gamma \rho}(\tau_2, 0)
R^{\delta \lambda}(\tau_1,- r')
\langle \Delta_{\alpha \beta; \delta} (u_{r,t_1} - u_{r, t_2})
\nonumber \\
&& [ 
\Delta_{\lambda \rho; \gamma}(u_{r+r', t_1-\tau_1} 
- u_{r+r', t_1-\tau_1-\tau_2}) -
\Delta_{\lambda \rho; \gamma}(u_{r+r', t_2-\tau_1} 
- u_{r+r', t_2-\tau_1-\tau_2}) ] \rangle_c
\end{eqnarray}

We have written these terms in that form for simplicity,
but one must keep in mind that in addition they
must be symmetrized under $\alpha \to \beta$ and
$r \to -r$ when necessary.
Note that, to avoid making the notations even more
heavy, we have deliberately chosen in this Appendix {\it not} to 
write systematically all space and time integrals and we
assume that the reader can restitute the necessary integrations.

We now identify the new operators generated in a 
short distance expansion. As discussed in the text we will not
pay special attention here to the kinetic term (i.e the friction
to second order as well as the contributions to the
time persistent kinetic term) but rather focus on the
contributions to the disorder. Note also that it is also
possible to use a supersymmetric perturbation theory formalism to study the
dynamics of this model. However since we may also be interested in the future
to cases which are {\it not} potential \cite{ledou_movglass}
and to which supersymmetry does not apply as conveniently,
we will use a more pedestrian (and painful) approach.

The above terms 
(\ref{terms}) give either a contribution of {\it fusion}
that we write under the form:

\begin{eqnarray} \label{definition}
\int_{r,t_1,t_2} \delta \Delta^{\alpha \beta}_K e^{i K (u_{r t_1} - u_{r t_2})}
e^{-\frac{1}{2} K.B_{0,t_1-t_2}.K }
\end{eqnarray}

( We have defined 
$B^{\alpha \beta}_{r,\tau}= 2 (C^{\alpha \beta}_{0,0} - C^{\alpha \beta}_{r,\tau})$,
identical to definition (\ref{definition-b}) and we use that
$B^{\alpha \beta}_{0,t_1-t_2+\tau_1} = B^{\alpha \beta}_{0,t_1-t_2}
+ O(\tau_1/(t_1-t_2))$ ),

or, they give a contribution {\it annihilation} of the form:
$- \frac{1}{2} (i \hat{u}^{\alpha}_{r t_1}) \nabla^2
(i \hat{u}^{\beta}_{r+r',t_2})$.
This analysis parallels the one of the preceding section
from a static replica calculation.

{\it (i) different $K$ fusion terms}

The terms which give (from the Coulomb gas analogy)
the fusion of the type $K_1+K_2 \to K_3$ are the last two terms of
$\delta \Delta^{(2)}$ and the first term of $\delta \Delta^{(3)}$.
The last two terms of $\delta \Delta^{(2)}$ give:

\begin{eqnarray}
\delta \Delta^{\alpha \beta}_{K+K'} =
\frac{1}{2} K_\gamma K_\delta \Delta^{\alpha \beta}_K \Delta^{\rho \lambda}_{K'}
R^{\gamma \rho}_{- r,\tau}
R^{\delta \lambda}_{-r,\tau'}
e^{K.(2 C_{0,0} - C_{r,- \tau} - C_{r,- \tau'} ).K'} 
\end{eqnarray}

where a factor $K B_{0,t_1-t_2} K'$ has been absorbed in front to make
$\exp( - 1/2 (K+K')^2 B_{0,t_1-t_2})$ appear in front, as required
by the definition (\ref{definition}) of $\Delta^{\alpha \beta}_{K+K'}$, and 

\begin{eqnarray}
\delta \Delta^{\alpha \beta}_{K-K'} =
\frac{1}{2} K_\gamma K_\delta \Delta^{\alpha \beta}_K \Delta^{\rho \lambda}_{K'}
R^{\gamma \rho}_{- r,\tau}
R^{\delta \lambda}_{-r,\tau'}
e^{- K.(2 C_{0,0} - C_{r,- \tau} - C_{r,- \tau'} ).K'} 
\end{eqnarray}
Adding to these terms the first term of $\delta \Delta^{(3)}$
gives a total contribution:

\begin{eqnarray}
\delta \Delta^{\alpha \beta}_{K+K'} =
\frac{1}{2} K_\gamma K_\delta \Delta^{\alpha \beta}_K ( \Delta^{\rho \lambda}_{K'}
+ \Delta^{\rho \lambda}_{-K'} )
R^{\gamma \rho}_{- r,\tau}
R^{\delta \lambda}_{-r,\tau'}
e^{K.(2 C_{0,0} - C_{r,- \tau} - C_{r,- \tau'} ).K'} 
+ \nonumber  \\
K'_\gamma K_\delta \Delta^{\alpha \rho}_K \Delta^{\beta \lambda}_{-K'}
R^{\gamma \rho}_{r,\tau}
R^{\delta \lambda}_{-r,\tau'}
e^{K.(2 C_{0,0} - C_{r,\tau} - C_{r,- \tau'} ).K'} 
\end{eqnarray}

Using now that $\Delta^{\alpha \beta}_K =  \Delta_K K_{\alpha} K_{\beta}$,
the symmetries $\Delta^{\alpha \beta}_{_K}=\Delta^{\alpha \beta}_K$,
the $r\to -r$ symmetry of $R$ and $C$, and the $\tau \to -\tau$
symmetry of $C$, one finds that all three terms combine into:

\begin{eqnarray}
\delta \Delta^{\alpha \beta}_{K+K'} = \Delta_K \Delta_{K'}
( K_\alpha K_\beta + K_\alpha K'_\beta )
(K.R_{r,\tau}.K') (K.R_{r,\tau'}.K')
e^{\frac{1}{2} K.(B_{r,\tau} + B_{r,\tau'} ).K'}
\end{eqnarray}

Putting together the contribution of $(K,K')$ and $(K',K)$ one has
finally:

\begin{eqnarray}
\delta \Delta^{\alpha \beta}_{K+K'} = \Delta_K \Delta_{K'}
(K+K')_\alpha (K+K')_\beta \int_{r,\tau,\tau'}
(K.R_{r,\tau}.K') (K.R_{r,\tau'}.K')
e^{\frac{1}{2} K.(B_{r,\tau} + B_{r,\tau'} ).K'}
\end{eqnarray}

This expression can be simplified using the
FDT theorem (\ref{fdt}) $\partial_t
C^{\alpha \beta}_{r,t} = T R^{\alpha \beta}_{r,t}$.
Integrating by part and noting that $B_{r,\tau=0}=B_{stat}(r)$
one finds:

\begin{eqnarray}
\delta \Delta^{\alpha \beta}_{K+K'} =  \Delta_K \Delta_{K'} \frac{1}{T^2}
(K+K')_\alpha (K+K')_\beta \int_{r}
e^{K.B_{r,0}.K'}
\end{eqnarray}

we have also assumed $K.B_{r,\tau=\infty}.K'=-\infty$. This is
yields the result (\ref{g-renorm}) if one uses $\Delta_K = g T^2$.
It coincides also with the result of the statics (see previous
section).

{\it (ii) annihilation terms}

The annihilation term (i.e the first term of $\delta \Delta^{(3)}$ 
for $K'=K$ 
and the last two terms of $\delta \Delta^{(2)}$
for $K'=- K$  ) gives (assuming the $K \to -K$ symmetry):

\begin{eqnarray}
&& \delta \Delta^{\alpha \beta}_{0}(r) =\sum_K \int_{\tau,\tau'} K_\gamma K_\delta 
( \delta(r) 
\Delta^{\alpha \beta}_K \Delta^{\rho \lambda}_{K} \int_{r'}
R^{\gamma \rho}_{- r',\tau}
R^{\delta \lambda}_{-r',\tau'}
e^{- K.(2 C_{0,0} - C_{r',- \tau} - C_{r',- \tau'} ).K} 
+ \nonumber  \\
&& \Delta^{\alpha \rho}_K \Delta^{\beta \lambda}_{K}
R^{\gamma \rho}_{r,\tau}
R^{\delta \lambda}_{-r,\tau'}
e^{- K.(2 C_{0,0} - C_{r,\tau} - C_{r,- \tau'} ).K}
\end{eqnarray}

Thus assuming all symmetries it yields:

\begin{eqnarray}
&& \delta \Delta^{\gamma \delta}_{0}(q) =
\int_r (e^{i q.r} -1) f^{\gamma \delta}(r) \sim 
- \frac{1}{2} q_{\alpha} q_{\beta}
\int_r r^{\alpha} r^{\beta} f^{\gamma \delta}(r)
\nonumber \\
&& f^{\alpha \beta}(r) =
 \sum_K \int_{\tau,\tau'} K_\gamma K_\delta
(\Delta_K)^2 (K. R_{r,\tau}.K) (K.R_{r,\tau'}.K)
e^{- K.(2 C_{0,0} - C_{r,\tau} - C_{r,- \tau'} ).K}
\end{eqnarray}

The time integrals can again be performed using FDT. One finds:

\begin{eqnarray}
f^{\gamma \delta}(r) = \frac{1}{T^2} \sum_K K_\gamma K_\delta (\Delta_K)^2 e^{- K.B_{r,0}.K}
\end{eqnarray}

Since, by definition $\delta \Delta^{\gamma \delta}_{0}(q) \sim
- T [ (\Delta_{11} - \Delta_{66}) q_{\gamma} q_{\delta}
+  \Delta_{66} q^2 \delta_{\gamma \delta} ] $, one gets:

\begin{eqnarray}
(\delta \Delta_{11} - \delta \Delta_{66}) \frac{1}{2} ( \delta_{\gamma \alpha} 
\delta_{\delta \beta} + \delta_{\gamma \beta} 
\delta_{\delta \alpha} ) +
\delta \Delta_{66} \delta_{\alpha \beta} \delta_{\gamma \delta}
=
\frac{1}{2} \int_r r^{\alpha} r^{\beta} f^{\gamma \delta}(r) 
\end{eqnarray}

which yields equations (\ref{d66-renorm},\ref{d11-renorm})
when using $\Delta_K = g T^2$.

{\it (iii) same $K$ fusion terms}

Finally the same $K$ fusion terms are the most delicate.
There are a priori {\it four} distinct contributions.
Let us start with the first one, which as we show vanishes
identically: the last two terms of $\delta \Delta^{(3)}$ give, respectively:

\begin{eqnarray}
\delta \Delta^{\alpha \beta}_K = K_\delta
\Delta^{\alpha \rho}_K \sum_{K'} K'_\gamma
\Delta^{\beta \lambda}_{K'}
R^{\gamma \rho}_{r,\tau_2}
R^{\delta \lambda}_{-r,\tau_1}
e^{-\frac{1}{2}  K'.B_{0,\tau_1 + \tau_2}.K' }
e^{- K.(C_{r,-\tau_1} - C_{r,\tau_2}).K'}
\end{eqnarray}

and

\begin{eqnarray}
\delta \Delta^{\alpha \beta}_{-K} = K_\gamma
\Delta^{\beta \lambda}_K \sum_{K'} K'_\delta
\Delta^{\alpha \rho}_{K'}
R^{\gamma \rho}_{r,\tau_2}
R^{\delta \lambda}_{-r,\tau_1}
e^{-\frac{1}{2}  K'.B_{0,\tau_1 + \tau_2}.K' }
e^{- K.(C_{r,\tau_2} - C_{r,-\tau_1}).K'}
\end{eqnarray}

However these last two terms give zero identically. Indeed
using all symmetries and definitions and remembering the
necessary symmetrization one finds first that
the last two terms of $\delta \Delta^{(3)}$ give:

\begin{eqnarray}
&& \delta \Delta^{\alpha \beta}_K = 
\sum_{K'} \Delta_K \Delta_{K'}
(K'.R_{r,\tau_2}.K)
(K'.R_{r,\tau_1}.K)
e^{-\frac{1}{2}  K'.B_{0,\tau_1 + \tau_2}.K' }
(K_\alpha K'_\beta -  K_\beta K'_\alpha) 
e^{- K.(C_{r,\tau_1} - C_{r,\tau_2}).K'}
\end{eqnarray}

Each term vanish because of
the symmetry $\tau_1 \to \tau_2$ which makes the summmand
over $K'$ odd under $K' \to -K'$.

The three remaining contributions are:

(i) The $\delta \Delta^{(1)}_{r'}$ term which gives:

\begin{eqnarray}
&& \delta \Delta^{\alpha \beta}_K =
- 2 \Delta^{\alpha \rho}_K \sum_{K'} K'_\gamma K'_\delta 
\Delta^{\beta \lambda}_{K'}
R^{\delta \lambda}_{0,\tau_2}
R^{\gamma \rho}_{r,\tau_1} 
e^{-\frac{1}{2}  K'.B_{0,\tau_2}.K' }
( e^{K.(C_{r,\tau_1} - C_{r,\tau_1-\tau_2}).K'} 
- e^{K.(C_{r,\tau_1+\tau_2} - C_{r,\tau_1}).K'}  )
\nonumber
\end{eqnarray}

(ii) the first two terms of $\delta \Delta^{(2)}$ which give
a contribution:

\begin{eqnarray}
\delta \Delta^{\alpha \beta}_K = - K_\gamma K_\delta
\Delta^{\alpha \beta}_K \sum_{K'}
\Delta^{\rho \lambda}_{K'}
R^{\gamma \rho}_{- r,\tau}
R^{\delta \lambda}_{-r,\tau'}
e^{-\frac{1}{2}  K'.B_{0,\tau' -\tau}.K' }
\cosh{K.(C_{r,- \tau} - C_{r,- \tau'}).K'}
\end{eqnarray}

(iii) the term  $\delta \Delta^{(4)}$ gives:

\begin{eqnarray}
\delta \Delta^{\alpha \beta}_{K} = K_\delta 
\Delta^{\alpha \beta}_K \sum_{K'} K'_\gamma
\Delta^{\lambda \rho}_{K'}
R^{\gamma \rho}_{0,\tau_2}
R^{\delta \lambda}_{-r,\tau_1}
e^{-\frac{1}{2}  K'.B_{0,\tau_2}.K' } ( 
e^{- K.(C_{r,-\tau_1} - C_{r,- \tau_1 - \tau_2}).K'} -
e^{K.(C_{r,-\tau_1} - C_{r,- \tau_1 - \tau_2}).K'} )
\nonumber 
\end{eqnarray}

Using FDT we will show that the last two contributions 
will almost cancel. First we need to rearrange the first
one. The $\delta \Delta^{(1)}_{r}$ term gives:

\begin{eqnarray}
&& \delta \Delta^{(1)}_K =
- \sum_{K'} ( K_\alpha K'_\beta + K'_\alpha K_\beta)
\Delta_K \Delta_{K'} (K'.R_{0,\tau_2}.K')
(K'.R_{r,\tau_1}.K) 
e^{-\frac{1}{2}  K'.B_{0,\tau_2}.K' } \nonumber \\
&&
( e^{K.(C_{r,\tau_1} - C_{r,\tau_1-\tau_2}).K'} 
- e^{K.(C_{r,\tau_1+\tau_2} - C_{r,\tau_1}).K'}  )
\nonumber
\end{eqnarray}

We now split this term in two parts:

\begin{eqnarray}
&& \delta \Delta^{(1,1)}_K =
- \sum_{K'} ( K_\alpha K'_\beta + K'_\alpha K_\beta)
\Delta_K \Delta_{K'} \int_{\tau_2 > \tau_1}
(K'.R_{0,\tau_2}.K')
(K'.R_{r,\tau_1}.K)
\nonumber \\
&&
e^{-\frac{1}{2}  K'.B_{0,\tau_2}.K' + K.(C_{r,\tau_1} - C_{r,\tau_2-\tau_1}).K'}
\nonumber
\end{eqnarray}

which can be changed into, using FDT:

\begin{eqnarray}
&& \delta \Delta^{(1,1)}_K =
- \sum_{K'} ( K_\alpha K'_\beta + K'_\alpha K_\beta)
\Delta_K \Delta_{K'} \int_{\tau_2 > \tau_1}
(K'.R_{r,\tau_2-\tau_1}.K)
(K'.R_{r,\tau_1}.K)
\nonumber \\
&&
e^{-\frac{1}{2}  K'.B_{0,\tau_2}.K' + K.(C_{r,\tau_1} - C_{r,\tau_2-\tau_1}).K'}
\nonumber \\
&&
+ (\frac{1}{T})
\sum_{K'} ( K_\alpha K'_\beta + K'_\alpha K_\beta)
\Delta_K \Delta_{K'} \int_{\tau_1}
(K'.R_{r,\tau_1}.K)
[ e^{-\frac{1}{2}  K'.B_{0,\tau_2}.K' 
+ K.(C_{r,\tau_1} - C_{r,\tau_2-\tau_1}).K'} ]_{
\tau_2=\tau_1}^{\infty}
\end{eqnarray}

The first integral can be rewritten using $\tau=\tau_1$ and
$\tau'=\tau_2-\tau_1$ and becomes symmetric in $\tau \to \tau'$.
It is still odd in $K'$ however and thus is identically zero.
Thus:

\begin{eqnarray}
&& \delta \Delta^{(1,1)}_K =
(\frac{1}{T})
\sum_{K'} ( K_\alpha K'_\beta + K'_\alpha K_\beta)
\Delta_K \Delta_{K'} \int_{\tau_1}
(K'.R_{r,\tau_1}.K)
[ e^{-\frac{1}{2}  K'.B_{0,\tau_2}.K' 
+ K.(C_{r,\tau_1} - C_{r,\tau_2-\tau_1}).K'} ]_{
\tau_2=\tau_1}^{\infty}
\end{eqnarray}

The second piece is:

\begin{eqnarray}
&& \delta \Delta^{(1,2)}_K =
- \sum_{K'} ( K_\alpha K'_\beta + K'_\alpha K_\beta)
\Delta_K \Delta_{K'} (K'.R_{0,\tau_2}.K')
(K'.R_{r,\tau_1}.K) 
e^{-\frac{1}{2}  K'.B_{0,\tau_2}.K' } \nonumber \\
&&
( e^{K.(C_{r,\tau_1} - C_{r,\tau_1-\tau_2}).K'} \theta(\tau_1-\tau_2)
- e^{K.(C_{r,\tau_1+\tau_2} - C_{r,\tau_1}).K'}  )
\nonumber
\end{eqnarray}

This last term can be transformed, changing $\tau_1 \to \tau_1 + \tau_2$
in the first integral, into:

\begin{eqnarray}
&& \delta \Delta^{(1,2)}_K =
- \sum_{K'} ( K_\alpha K'_\beta + K'_\alpha K_\beta)
\Delta_K \Delta_{K'} (K'.R_{0,\tau_2}.K')
(K'.R_{r,\tau_1+\tau_2 }.K - K'.R_{r,\tau_1}.K)
\nonumber \\
&&
e^{-\frac{1}{2}  K'.B_{0,\tau_2}.K' } 
e^{K.(C_{r,\tau_1+\tau_2} - C_{r,\tau_1}).K'}
\end{eqnarray}

which can be integrated using FDT into:

\begin{eqnarray}
\delta \Delta^{(1,2)}_K =
(\frac{1}{T}) \sum_{K'} ( K_\alpha K'_\beta + K'_\alpha K_\beta)
\Delta_K \Delta_{K'} \int_{\tau_2}
(K'.R_{0,\tau_2}.K')
e^{-\frac{1}{2}  K'.B_{0,\tau_2}.K' }
[e^{K.(C_{r,\tau_1+\tau_2} - C_{r,\tau_1}).K'}]_{
\tau_1=0}^{\tau_1=\infty}
\nonumber
\end{eqnarray}

Similarly the first two terms of $\delta \Delta^{(2)}$ give
a total contribution:

\begin{eqnarray}
\delta \Delta^{(2)}_K = 
- K_\alpha K_\beta \sum_{K'} \Delta_K \Delta_{K'} 
(K.R_{r,\tau}.K')
(K.R_{r,\tau'}.K')
e^{-\frac{1}{2}  K'.B_{0,\tau' -\tau}.K' }
e^{K.(C_{r,\tau} - C_{r,\tau'}).K'}
\end{eqnarray}

The term  $\delta \Delta^{(4)}$ gives:

\begin{eqnarray}
&& \delta \Delta^{(4)}_K = 
2 K_\alpha K_\beta \sum_{K'} \Delta_K \Delta_{K'}
(K'.R_{0,\tau_2}.K')
(K.R_{r,\tau_1}.K')
e^{-\frac{1}{2}  K'.B_{0,\tau_2}.K' } e^{- K.(C_{r,\tau_1} - C_{r,\tau_1 + \tau_2}).K'} 
\nonumber 
\end{eqnarray}

Using FDT this term and integrating by part over $\tau_2$
this term gives:

\begin{eqnarray}
&& \delta \Delta^{(4)}_K = 
- 2 K_\alpha K_\beta \sum_{K'} \Delta_K \Delta_{K'}
(K'.R_{r,\tau_1 + \tau_2}.K)
(K.R_{r,\tau_1}.K')
e^{-\frac{1}{2}  K'.B_{0,\tau_2}.K' }
e^{- K.(C_{r,\tau_1} - C_{r,\tau_1 + \tau_2}).K'}
- \nonumber \\
&&
(\frac{-1}{T})
2 K_\alpha K_\beta \sum_{K'} \Delta_K \Delta_{K'}
\int_{\tau_1} (K.R_{r,\tau_1}.K') 
[ e^{-\frac{1}{2}  K'.B_{0,\tau_2}.K' 
-
K.(C_{r,\tau_1} - C_{r,\tau_1 + \tau_2}).K'} ]_{\tau_2=0}^{\tau_2=\infty}
\nonumber 
\end{eqnarray}

and thus:

\begin{eqnarray}
&& \delta \Delta^{(4)}_K =
- \delta \Delta^{(2)} + 
(\frac{1}{T})
2 K_\alpha K_\beta \sum_{K'} \Delta_K \Delta_{K'}
\int_{\tau_1} (K.R_{r,\tau_1}.K') 
[ e^{-\frac{1}{2}  K'.B_{0,\tau_2}.K' 
-
K.(C_{r,\tau_1} - C_{r,\tau_1 + \tau_2}).K'} ]_{\tau_2=0}^{\tau_2=\infty}
\end{eqnarray}

where $\tau=\tau_1+\tau_2$ and $\tau'=\tau_1$ has been used,
as well as the symmetry in $\tau \to \tau'$ of 
$\delta \Delta^{(2)}$ which allows to write it as 
twice the integral over $\tau \ge \tau'$.

Thus, one has:

\begin{eqnarray}
&& \delta \Delta^{(1)}_K + \delta \Delta^{(2)}_K +
\delta \Delta^{(4)}_K =
(\frac{1}{T})
\sum_{K'} ( K_\alpha K'_\beta + K'_\alpha K_\beta)
\Delta_K \Delta_{K'} \nonumber \\
&&
( 
\int_{\tau_1}
(K'.R_{r,\tau_1}.K)
[ e^{-\frac{1}{2}  K'.B_{0,\tau_2}.K' 
+ K.(C_{r,\tau_1} - C_{r,\tau_2-\tau_1}).K'} ]_{
\tau_2=\tau_1}^{\tau_2=\infty} +
\nonumber \\
&&
\int_{\tau_2}
(K'.R_{0,\tau_2}.K')
e^{-\frac{1}{2}  K'.B_{0,\tau_2}.K' }
[e^{K.(C_{r,\tau_1+\tau_2} - C_{r,\tau_1}).K'}]_{
\tau_1=0}^{\tau_1=\infty}
)
+ \nonumber \\
&& 
(\frac{1}{T})
2 K_\alpha K_\beta \sum_{K'} \Delta_K \Delta_{K'}
\int_{\tau_1} (K.R_{r,\tau_1}.K') 
[ e^{-\frac{1}{2}  K'.B_{0,\tau_2}.K' 
-
K.(C_{r,\tau_1} - C_{r,\tau_1 + \tau_2}).K'} ]_{\tau_2=0}^{\tau_2=\infty}
\end{eqnarray}

This can be rewritten as:

\begin{eqnarray}
&& \delta \Delta^{(1)}_K + \delta \Delta^{(2)}_K +
\delta \Delta^{(4)}_K =
(\frac{1}{T})
\sum_{K'} ( K_\alpha K'_\beta + K'_\alpha K_\beta)
\Delta_K \Delta_{K'} \nonumber \\
&&
\int_{\tau} 
(K'.R_{r,\tau}.K)
e^{-K'.C_{0,0}.K' + K.C_{r,\tau}.K'}
+ (K'.R_{0,\tau}.K')
e^{-K'.C_{0,0}.K' + K'.C_{0,\tau}.K'} 
\nonumber \\
&&
- \int_{\tau} 
(K'.R_{r,\tau}.K + K'.R_{0,\tau}.K')
e^{-K'.(C_{0,0}- C_{0,\tau}).K' 
+ K.(C_{r,\tau}-C_{r,0}).K'}
\nonumber \\
&&
+ (\frac{2}{T})
K_\alpha K_\beta \sum_{K'} \Delta_K \Delta_{K'}
\int_{\tau}
(K'.R_{r,\tau}.K)
( e^{-K'.C_{0,0}.K' - K'.C_{r,\tau}).K}
-1 )
\end{eqnarray}

Using the symmetry $K' \to -K'$ one see that the
second term vanishes. Using FDT and this symmetry one
finds that the first and third term cancel. Finally
one finds:

\begin{eqnarray}
&& \delta \Delta^{(1)}_K + \delta \Delta^{(2)}_K +
\delta \Delta^{(4)}_K =
- (\frac{2}{T^2}) 
K_\alpha K_\beta \sum_{K'} \Delta_K \Delta_{K'}
\int_{\tau}
e^{-K'.C_{0,0}.K' - K'.C_{r,0}.K}
\end{eqnarray}

The divergent integrals at $T_g$ correspond to 
$K' = -K$, with the 
final result:

\begin{eqnarray}
&& \delta \Delta^{(1)}_K + \delta \Delta^{(2)}_K +
\delta \Delta^{(4)}_K =
- (\frac{2}{T^2}) 
K_\alpha K_\beta \sum_{K'} (\Delta_K)^2 
\int_{\tau,r}
e^{-K.(C_{0,0} - C_{r,0}).K}
\end{eqnarray}

Note that we have used that
$C_{r,\tau=\infty}=0$.

\deuxcol

\section{Statistical symmetry and correlation function}
\label{appendixe}

In this Appendix we detail some useful consequences of the statistical tilt symmetry.
It allows for instance to obtain a useful form for
the real space Callan-Symanzik equations referred to in the text.
We consider for simplicity the $N=1$ planar random 
field XY model defined in its replicated version as (\ref{cardyostlund}), the
generalisation to (\ref{def-model}) being straightforward.

First we derive the Ward Identities of the statistical tilt symmetry
on the replicated model. Under a local tilt 
$u^a(x) \rightarrow u^a(x)+\epsilon(x)$ the correlation generating functional $Z^{(n)}[J]$ remains 
unchanged. This invariance can be written 
$$ \int d^2 x \epsilon(x) \left( \frac{c}{T} \nabla^2 \sum_a  \frac{\delta}{\delta J^{a}(x)}+ 
\sum_a J_a(x) \right) Z^{(n)}[J] =0$$
 Introducing the generating functional of the connected correlation $W^{(n)}[J]$ and 
using a Legendre transform \cite{zinn-justin2} we arrive at the Ward identities for the 
1PI replicated functional $\Gamma^{(n)}[u]$ :
\begin{equation}
- \frac{c}{T} \sum_a \nabla^2 u^a(x) = \sum_a \frac{\delta \Gamma^{(n)}[u]}{\delta u^a(x)}
\end{equation}
>From this one deduces useful consequences, e.g that $c$ is not renormalized to all order. 

Another useful consequence of this symmetry is to relate the
correlation functions of the original model to the one
of the model with $\Delta=0$, as noted in \cite{giamarchi-vortex-long}.
This is easier to demonstrate on
the unreplicated model (analogous to (\ref{def-model})).
Consider:

\begin{equation} \label{correl-stat-sym}
G_h(x)=\overline{ \langle e^{i \int h(x) u(x)} \rangle}
\end{equation}

Using the local shift
$ \nabla u\rightarrow \nabla u -\frac{T}{c}  \vec{\sigma}$ where 
$\vec{\sigma}$ is the local random stress for the $N=1$ version
of model (\ref{def-model}) (one can assume $\vec{\sigma}_q$ to
be purely longitudinal), we can (up to a redefinition 
of the random field $\phi$ which does not modify the
average ) factorize out the $\Delta$ dependance of the correlation function 
(\ref{correl-stat-sym}) :

\begin{equation} \label{correl-stat-sym-2}
G_h(x)= e^{ -  \frac{1}{2} \int_q \frac{T \Delta}{c^2 q^2}  h_q h_{-q} } ~~ G_h(x)|_{\Delta=0}  
\end{equation}

If one consider for instance the regularized two point correlation function of the 
operator $e^{ipu(x)}$:

\begin{equation} \label{correl-stat-sym-3}
G(x,a)=\overline{ \langle e^{i pu(x) }e^{-i pu(0) } \rangle  }
\end{equation}

obtained from the above with for $h(y) = p \delta(y-x) - p \delta(y)$, 
one has:

\begin{equation}
G(r,a)= Z(\Delta,T) .\left. G(r,a) \right|_{\Delta=0}
\end{equation}

where $\left. G(r,a) \right|_{\Delta=0}$ 
is the correlation function with $\Delta$ set to zero in the model (),
and the factor 
\begin{eqnarray}
\ln Z (\Delta,T)= - \frac{p^2 T \Delta}{c^2} 
 \int \frac{d^2q}{(2\pi)^2}\frac{1 - \cos(q.r)}{ q^2}
\end{eqnarray}
 Using formula (\ref{formul-int-ang}) we find 
\begin{equation}
\ln Z (\Delta,T)= - \frac{p^2 T \Delta}{2 \pi c^2} \ln \left( \frac{r}{a} \right)
\end{equation}

This can now be used to study the real space Callan-Symanzik equation \cite{denis-bernard}.
The two point function behaves under a rescaling as:
\begin{equation}
G(r,\Delta(0),\tilde{g}(0))=e^{-l d_p^*} G(r e^{-l},\Delta(l),\tilde{g}(l))
\end{equation}
 where $d_p^*$ is the anomalous dimension of the operator $e^{i p u(x)}$
(which was computed e.g in Appendix B of \cite{toner-log-2}).
The invariance of this correlation function 
under a change of cut-off parameter $a\rightarrow ae^l$ gives the 
Callan-Symanzik equation:
\begin{equation}\label{callanS}
\left(
-\frac{\partial}{\partial \ln r} - d_p^* + \frac{d\Delta }{dl} \frac{\partial}{\partial \Delta}
+   \frac{d \tilde{g} }{dl} \frac{\partial}{\partial  \tilde{g}} \right)G(r,a)=0
\end{equation}

Using the above relation into (\ref{callanS}) we obtain 
$$\left(
-\frac{\partial}{\partial \ln r} - d_p^* -\frac{p^2 T}{2 \pi c^2} \ln \left( \frac{r}{a} \right)
   \frac{d\Delta(l)}{dl}
+   \frac{d \tilde{g} }{dl} \frac{\partial}{\partial  \tilde{g}} \right)G(r,a)=0
$$
 Thus we obtain that 
if the operator $e^{ipu(x)}$ is multiplicatively renormalisable to all order, 
the dominant term in the large $r$ limit will be to all order an 
$\exp ( \lambda.\ln^2(r/a))$ term. To first order, using (\ref{RG-co}) we find on 
the low temperature fixed line ($d\tilde{g}/dl =0$), near $T_g=4\pi c, \tilde{g}=\tau/2\pi$ :
$$ \ln G(r,a) = -p^2 \tau^2 \ln^2\left( \frac{r}{a} \right) + {\cal O}(\ln (r/a))$$

Note that this shows that the dominant term ($r\gg a$) has {\it a gaussian distribution}.
Thus one obtains permuting the  
limit $p\rightarrow 0$ and the disorder average):
\begin{eqnarray} \label{result-structure}
B(r) = - 2 \frac{1}{p^2} \ln G(r,a)= 2 \tau^2 \ln^2 (r/a) + {\cal O}(\ln (r/a))
\end{eqnarray}

which is the result obtained in Section (\ref{section3b2}) by another method.

\unecol


\begin{thebibliography}{999}
\bibitem{footnote3}
Note that since $T_g$ is quite an elevated temperature
the thermal displacement is not very small. One finds
$\langle \nabla u \rangle_T  \sim 0.15$ near $T_g$.
Thus even if dislocations
are excluded by hand, one is not in a controllable
way in the elastic limit as far as thermal displacements
are concerned (though extra disorder induced diplacements
can be made as small as possible). Thus there is still
an (natural) assumption in the present study,
namely that the original lattice
model is within the basin of attraction of
the fixed point studied (i.e of model
(\ref{def-model})). It would be
interesting to check numerically whether this 
assumption is correct.
\bibitem{footnote6}
 Note that one can consider the exact interaction $V_{ij}(r)$ instead of its 
large $r$ form through the Coulomb Gas approach. Eq. (\ref{perturb-elastic}) 
remains valid and one has to develop the $\tilde{g}^3$ term, keeping only the 
relevant operators. Using the results of appendix A, we recover exactly the 
RG eqs. of the effective action method (appendix B).
\bibitem{footnote}
It is known that renormalisation on the correlation function is equivalent to renormalisation on 
the partition function for the KT transition\cite{nelson-in-domb}. Moreover, we have checked that we recover 
with this scheme 
the results of Cardy and Ostlund for the random field XY model without dislocations\cite{cardy-desordre-rg}.
\bibitem{footnote2} Analytical solutions of mean field dynamics
\cite{meanfield-dynamics1,meanfield-dynamics2}
suggests that the replica symmetry breaking RG constructed
in the statics
in Ref \cite{ledoussal-rsb-lettre})
could be extended to nonequilibrium dynamics.
For instance, the relation 
$\partial_{t'} D^{\alpha \beta}(u_{rt} - u_{rt'}, t,t')
= T \Sigma^{\alpha \beta}(u_{rt} - u_{rt'}, t,t')$ could be 
generalized to a nonequilibrium perturbation theory
as a new set of Ward Identities 
$\partial_{t'} D^{\alpha \beta}(u_{rt} - u_{rt'}, t,t')
= X \Sigma^{\alpha \beta}(u_{rt} - u_{rt'}, t,t')$
for an arbitrary generalized FDT ratio.
\end{thebibliography}
\end{document}